\documentclass[%
preprint,
superscriptaddress,
nofootinbib,
amsmath,amssymb,
longbibliography,
]{revtex4-1}
\usepackage{lipsum}
\usepackage{amsmath}
\usepackage{amssymb}
\usepackage{graphicx}
\usepackage{braket}
\usepackage{color}
\usepackage{xcolor}
\usepackage{siunitx}
\usepackage{upgreek}
\usepackage[utf8]{inputenc}
\usepackage{physics}
\usepackage{multirow}
\usepackage{array}
\usepackage{url}
\usepackage{dcolumn}
\usepackage{bm}
\usepackage{hyperref}
\usepackage{natbib}
\usepackage{gensymb}
\usepackage{placeins}
\usepackage{setspace}
\usepackage[bottom]{footmisc}

\begin{document}


\title{A single spin in hexagonal boron nitride for vectorial quantum magnetometry}

\author{Carmem~M.~Gilardoni}
\thanks{These authors contributed equally to this work} 
\affiliation{Cavendish Laboratory, JJ Thomson Avenue, University of Cambridge, Cambridge CB3 0HE, UK}
\author{Simone~Eizagirre~Barker}
\thanks{These authors contributed equally to this work} 
\affiliation{Cavendish Laboratory, JJ Thomson Avenue, University of Cambridge, Cambridge CB3 0HE, UK}
\author{Catherine~L.~Curtin}
\affiliation{Cavendish Laboratory, JJ Thomson Avenue, University of Cambridge, Cambridge CB3 0HE, UK}
\author{Stephanie~A.~Fraser}
\affiliation{Cavendish Laboratory, JJ Thomson Avenue, University of Cambridge, Cambridge CB3 0HE, UK}
\author{Oliver.~F.J.~Powell}
\affiliation{Cavendish Laboratory, JJ Thomson Avenue, University of Cambridge, Cambridge CB3 0HE, UK}
\affiliation{Hitachi Cambridge Laboratory, Hitachi Europe Ltd., JJ Thomson Avenue, Cambridge CB3 0HE, UK}
\author{Dillon~K.~Lewis}
\affiliation{Cavendish Laboratory, JJ Thomson Avenue, University of Cambridge, Cambridge CB3 0HE, UK}
\author{Xiaoxi~Deng}
\affiliation{Cavendish Laboratory, JJ Thomson Avenue, University of Cambridge, Cambridge CB3 0HE, UK}
\author{Andrew~J.~Ramsay}
\affiliation{Hitachi Cambridge Laboratory, Hitachi Europe Ltd., JJ Thomson Avenue, Cambridge CB3 0HE, UK}
\author{Chi~Li}
\affiliation{ARC Centre of Excellence for Transformative Meta-Optical Systems, Faculty of Science, University of Technology Sydney, Ultimo, New South Wales, Australia}
\affiliation{School of Mathematical and Physical Sciences, Faculty of Science, University of Technology Sydney, Ultimo, New South Wales, Australia}
\author{Igor~Aharonovich}
\affiliation{ARC Centre of Excellence for Transformative Meta-Optical Systems, Faculty of Science, University of Technology Sydney, Ultimo, New South Wales, Australia}
\affiliation{School of Mathematical and Physical Sciences, Faculty of Science, University of Technology Sydney, Ultimo, New South Wales, Australia}
\author{Hark~Hoe~Tan}
\affiliation{ARC Centre of Excellence for Transformative Meta-Optical Systems, Department of Electronic Materials Engineering, Research School of Physics, The Australian National University, Canberra, ACT 2600, Australia}
\author{Mete~Atat\"ure}
\affiliation{Cavendish Laboratory, JJ Thomson Avenue, University of Cambridge, Cambridge CB3 0HE, UK}
\author{Hannah~L.~Stern}
\affiliation{Photon Science Institute, Department of Physics and Department of Chemistry, The University of Manchester, Manchester, M13 9PL, UK}

\date{Version of \today}

\begin{abstract}

Quantum sensing based on solid-state spin defects provides a uniquely versatile platform for imaging physical properties at the nanoscale under diverse environmental conditions. 
Operation of most sensors used to-date is based on projective measurement along a single axis combined with computational extrapolation.
Here, we show that the individually addressable carbon-related spin defect in hexagonal boron nitride is a multi-axis spin system for vectorial nanoscale magnetometry.
We demonstrate how its low symmetry and strongly spin-selective direct and reverse intersystem crossing dynamics provide sub-$\mu\text{T}/\sqrt{\text{Hz}}$ magnetic-field sensitivity for both on and off-axis bias magnetic field exceeding 50~mT. 
Alongside these features, the room-temperature operation and the nanometer-scale proximity enabled by the van der Waals host material further consolidate this system as an exciting quantum sensing platform. 

\vspace{35mm} 

\end{abstract}

\let\thefootnote\relax\footnotetext{Correspondence to: CMG (cm2207@cam.ac.uk) or HLS (hannah.stern@manchester.ac.uk)}

\maketitle


\textbf{Introduction}

Spin defects in solids can be used as quantum sensors to study phenomena across condensed matter, geological and biological systems \cite{Budker2007,Degen2017,Glenn2017}. When reduced to the single spin level, optically addressable high-spin (\textit{S} \textgreater 1/2) defects can provide quantitative field, temperature and pressure sensors with nanoscale spatial resolution, in a technique described as nanoscale quantum sensing \cite{Chernobrod2005}. The rapid development of quantum sensors has been driven largely by the nitrogen-vacancy (NV) centre in diamond \cite{Taylor2008,Balasubramanian2008,Degen2008,Maze2008,Rondin2012,Rondin2014, Barry2020,Rovny2024}, as well as defects in silicon carbide \cite{Simin,Lee2015,Niethammer2016}. 
For DC sensing, pioneering work with the NV centre has demonstrated mapping of static fields formed by spin order and current flow in materials \cite{Maletinsky2012,Tetienne2012,Pelliccione2016,Chang2017,Dovzhenko2018,Appel2019,Haykal2020,Finco2021}, including in atomically thin semiconductors \cite{Thiel2019,Sun2021}. Much of this work has provided key fundamental insight into the nature of magnetisation in these materials \cite{Tetienne2015,Tan2023}. In addition, the NV centre can probe local properties inside cells, including temperature and magnetic fields \cite{LeSage2013,Wu2016,Davis2018}. \\

An outstanding challenge for magnetometry using uniaxial spin systems, like the NV centre, is that the vector target field cannot be unambiguously determined, because the spin transitions are only sensitive to the projection of external field along the high-symmetry axis \cite{Rondin2014,Lee2015}. 
In addition, NV-based magnetometry faces challenges regarding its operation under arbitrarily oriented magnetic fields that exceed $\sim$10 mT, conditions that are often met when studying ferromagnetic nanostructures \cite{Rondin2014}. This is because strong off-axis magnetic fields lead to spin mixing that degrade the optical initialisation process \cite{Tetienne2012}. 
In NV-based experiments where the target materials generate large off-axis magnetic field, readout can be preserved by increasing the target-NV distance, thereby limiting the spatial resolution.
A high-spin defect that can overcome the limitations of dynamic range while providing full vectorial sensitivity and \textless 10~nm spatial resolution, with operation over a broad temperature range, would dramatically increase the throughput and the scope of nanoscale quantum sensing. \\

Spin defects in two-dimensional materials offer an alternative platform for realising quantum-sensor-based microscopy, where the atomic thickness and layered nature of the host material offers the potential to achieve higher spatial resolution and provide new opportunities for integration into hybrid devices \cite{Novoselov2016,Liu2019}. Wide-field quantum microscopy with the \textit{S}=1 optically addressable boron vacancy ($\textit{V}_\text{B}^-$) spin defect ensembles in hexagonal boron nitride (hBN) show the versatility of the hBN host material regarding integration with 2D heterostructures \cite{Huang2022,Healey2022,Vaidya2023}. These reports present mapping of magnetic domains, temperature and charge currents in layered ferromagnets, albeit with diffraction-limited spatial resolution due to use of defect ensembles. \\

In this article, we reveal that the \textit{S}=1 carbon-related spin defect in hBN \cite{Stern2022,Stern2024} presents an attractive system for nanoscale quantum sensing, displaying both high DC sensitivity, broad magnetic field dynamic range and potential for unprecedented spatial resolution. We show that, due to its low-symmetry and advantageous excited state dynamics, this spin defect displays multiple ground-state optically detected magnetic resonances (ODMR) with contrast that can exceed 90\% (see Supplementary Materials, Fig.~S1) and a signal that persists at arbitrarily orientated magnetic field beyond 50~mT. We combine photon-emission correlation spectroscopy and pulsed ODMR experiments with microscopic modeling of the optical cycle to explain the kinetic origin of the high ODMR contrast and large dynamic range for this quantum sensor. Finally, we show that this system is capable of achieving DC sensitivity reaching 500~nT/$\sqrt{\text{Hz}}$, even in the presence of strong bias field, a region that is currently out of range for NV-based nanoscale magnetometry \cite{Tetienne2012}. We show that full vectorial mapping of a target field is accessible via dual-axis readout of multiple spin resonances for a single hBN defect. Our results present a new candidate for nanoscale quantum sensing that has the potential to extend the technique to new systems across condensed matter physics, biology and biomedical science. \\



\textbf{An \textit{S}=1 system with dynamic range at high magnetic field}

\begin{figure*}[t]
  \includegraphics[width=\textwidth]{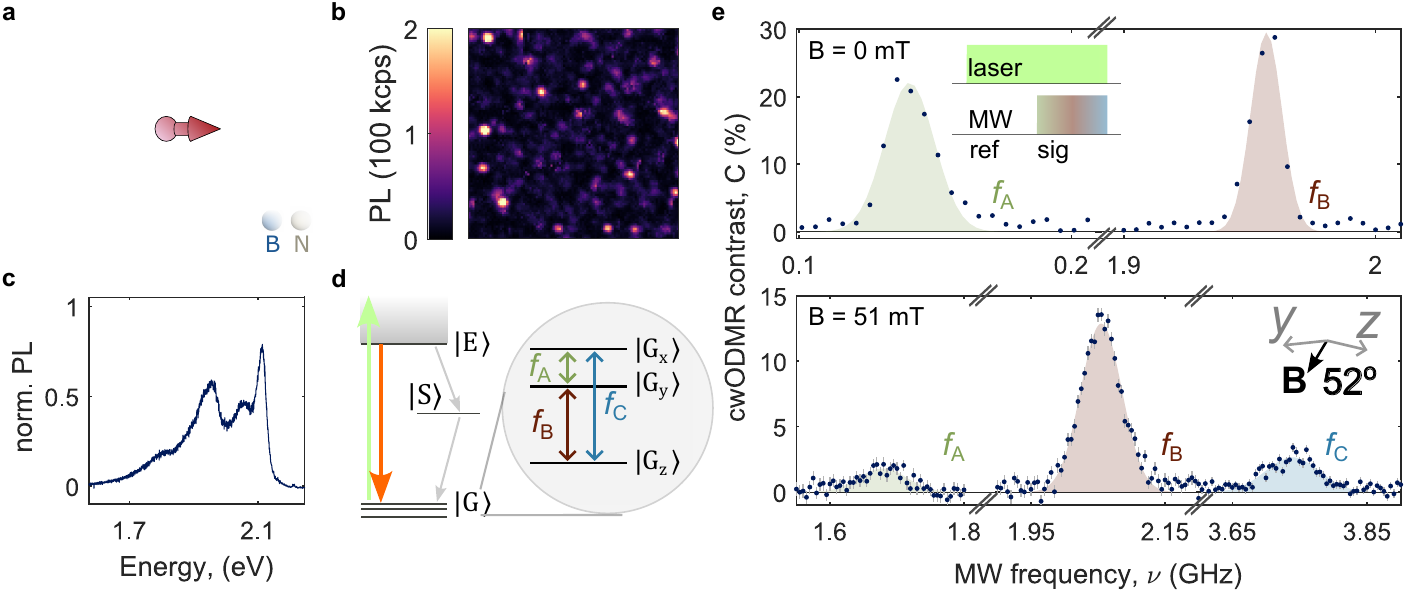}
  \caption{\textbf{ODMR persistence with applied magnetic field.} \textbf{a} Schematic of the hBN layers containing a spin defect with in-plane spin. \textbf{b} PL map and \textbf{c} photoluminescence spectrum of the carbon-related defect in hBN. \textbf{d} Schematic of the electronic level structure of the defects, consisting of ground and optically excited state manifolds, and a metastable state. Relaxation from the optically excited state to the ground-state manifold can occur radiatively or non-radiatively through a sequence of direct and reverse intersystem crossing events. The ground-state manifold is a spin-1 with non-degenerate spin sublevels at zero magnetic field. Spin resonance transitions between each of the three spin sublevels are possible, giving rise to three spin-resonance signatures, \textit{f}\textsubscript{A,B,C} in ascending energy. \textbf{e} cwODMR spectra measured at 0~mT (top panel) and 51(1)~mT (bottom panel), showing three possible spin transitions between the spin sublevels of an S=1 system. Blue circles are measured mean values, with gray error bars indicating standard error of mean. Shaded regions are fits to the data using a Gaussian peakshape. The inset in the top panel presents the pulse sequence used for detecting cwODMR, whereas the inset in the bottom panel present the direction of the magnetic field with respect to the defect's symmetry axes. 
  \label{Fig::fig1}}
\end{figure*}

Figure ~\ref{Fig::fig1} presents the carbon-related hBN spin defects investigated in this work, represented schematically in Fig.~\ref{Fig::fig1}a. This defect is grown into wafer-scale multilayer (30~nm thick) hBN via metal organic vapour phase epitaxy (MOVPE) in the presence of triethylboron \cite{Chugh2018, Mendelson2021,Stern2022,Stern2024}. This results in individually addressable, bright spin defects (typical saturation count rates in the range 5-200 kcps) that are resolved via scanning confocal microscopy with 532~nm illumination (see Fig.~\ref{Fig::fig1}b). 
Figure~\ref{Fig::fig1}c presents an example photoluminescence (PL) spectrum measured at room temperature, showing zero-phonon line emission at $\sim$2.1~eV accompanied by lower energy phonon side band typical of visible hBN defects \cite{Tran2016,Jungwirth2017,Grosso2017,Mendelson2021,Stern2022}. 
Figure~\ref{Fig::fig1}d presents the defect electronic structure, with spin-triplet ground and optically excited states and a spin-singlet metastable state \cite{Stern2024}. Relaxation from the optically excited state to the ground-state manifold can occur radiatively or non-radiatively through a sequence of spin-dependent direct and reverse intersystem crossing events that are responsible for optical spin initialisation. The ground-state spin triplet gives rise to three possible paramagnetic transitions between the three spin sublevels, labelled \textit{f}\textsubscript{A-C}. We study the properties of the ground-state spin via optically detected magnetic resonance (ODMR). Our experimental setup consists of a home-built confocal microscope equipped with a permanent magnet that can be moved in proximity and orientation with respect to the device, enabling magnetic field up to 140~mT. A coil in the vicinity of the device delivers microwaves to the hBN defect \cite{Stern2024,Stern2022}.\\

Fig.~\ref{Fig::fig1}e (top panel) shows the room-temperature ODMR spectrum for an hBN defect at 0~mT, where the microwaves were applied in the range 0.01 - 3 GHz. The inset shows the measurement sequence for detecting the continuous wave (cw) ODMR contrast, defined as the relative change in photoluminescence (PL) under 532-nm illumination induced by the presence of microwaves ($C = (\text{PL}_{\text{sig}}-\text{PL}_{\text{ref}})/\text{PL}_{\text{ref}}$). For this defect, we observe two ODMR resonances, at 0.129 and 1.95 GHZ, with comparable saturated cwODMR contrast of 22(5)\% and 30(2)\%, (see Fig. S3 for zero-field contrast statistics for a range of defects) \cite{Stern2024}. 
We assign the ODMR resonances to the transitions of the \textit{S}=1 system based on a Hamiltonian of the type,

\begin{align}
    H &= H\textsubscript{ZF} + H\textsubscript{ZE},
    \label{Eq::HamiltonianI}\\
    H\textsubscript{ZF} &= D S_z^2 + E (S_x^2 - S_y^2), \label{Eq::HamiltonianII} \\
    H\textsubscript{Ze} &= \gamma_e \mathbf{B} \cdot \mathbf{S},
    \label{Eq::HamiltonianIII}
\end{align}

\noindent where $H_{\text{ZF}}$ is the zero-field splitting term, $H_{\text{ZE}}$ is the Zeeman term, $D$ and $E$ are the zero-field splitting (ZFS) parameters that define the defect's $x,y,z$ principal axes in units of Hz, $\mathbf{S}$ is the \textit{S}=1 operator, $\gamma_e$ is the electron gyromagnetic ratio and $\mathbf{B}$ is the applied magnetic field. 
In the absence of applied magnetic field, we only need to consider the $H_{\text{ZF}}$ term with eigenenergies $0$, $D-E$, and $D+E$. \\

The magnitude of the transverse zero-field splitting $|E|$ is a measure of the rhombicity, or low symmetry, of the spin density of the system \cite{Richert2017}. In systems where $|E|$ is low compared to the linewidth (ie. for the NV centre in diamond and the $\textit{V}_{B}^{-}$ defect in hBN), overlapping resonances are observed at zero field, corresponding to transitions between $\ket{m_s=0}$ and the near-degenerate $\ket{m_s =\pm 1}$, where $m_s$ denotes the spin projection along the defect's $z$ axis \cite{Poole1974}. In such systems, the spin transitions give partial information of the vector of external magnetic field -- while the projection of the field along the $z$ axis (polar dependence) can be determined, the azimuthal direction cannot. 
In contrast, in the case of low-symmetry \textit{S}=1 systems, where $|E|$ $\neq$ 0, three transitions may arise between the three spin sublevels indicated in Fig.~\ref{Fig::fig1}c \cite{kohler,Wrachtrup1993,Lee2013,Richert2017,Foglszinger2022}. 
In this case, the transverse zero-field splitting term $E(S_x^2 - S_y^2)$ hybridises $\ket{m_s =\pm 1}$, relaxing the selection rules for transitions between them \cite{Wrachtrup1993}. 
The zero-field spin eigenstates are then given by $\ket{\text{G\textsubscript{z}}} = \ket{m_s=0}$, $\ket{\text{G\textsubscript{x}}} = (\ket{m_s=+1}-\ket{m_s=-1})/\sqrt{2}$, and $\ket{\text{G\textsubscript{y}}} = (\ket{m_s=+1}+\ket{m_s=-1})/\sqrt{2}$.
We assign the zero-field resonances shown in Fig.~\ref{Fig::fig1}e (top) to the transition between $\ket{\text{G\textsubscript{x}}}$ and $\ket{\text{G\textsubscript{y}}}$ (\textit{f}\textsubscript{A}), and $\ket{\text{G\textsubscript{z}}}$ and  $\ket{\text{G\textsubscript{y}}}$ (\textit{f}\textsubscript{B}), where $|D|$ = 2.025 GHz and $|E|$ = 70~MHz for this defect. 
Previous work on this defect type has reported the presence of all three transitions, but \textit{f}\textsubscript{A} was outside of the studied measurement range at zero field \cite{Stern2024}. \\

Fig.~\ref{Fig::fig1}e (bottom panel) presents the ODMR spectrum for the same defect under 51~mT magnetic field applied in the plane of the hBN layers. 
At this field, all three spin transitions are visible in the spectrum, with $C$(\textit{f}\textsubscript{A})=$1.8(2)\%$, $C$(\textit{f}\textsubscript{B})=$12.9(5)\%$, and $C$(\textit{f}\textsubscript{C})=$2.7(3)\%$, where $C$(\textit{f}\textsubscript{i}) is the contrast of the i-th transition. 
We determine the field vector is at $\ang{51(1)}$ from the defect $z$-axis, parallel to the $yz$ plane, from field-dependent measurements. 
This means that the defect's $y$ and $z$ axes are parallel to the plane of the hBN layers.   
Despite the high off-axis applied field, we observe that the ODMR resonance is not quenched.
This is in stark contrast to what is seen for the NV centre, where a magnetic field $\sim$10~mT misaligned to the defect's quantization axis quenches the ODMR resonances due to degradation of the spin initialisation mechanism \cite{Tetienne2012,Doherty2013}. \\


\textbf{Photodynamics of the carbon-related hBN spin}

\begin{figure*}[t]
  \includegraphics[width=\textwidth]{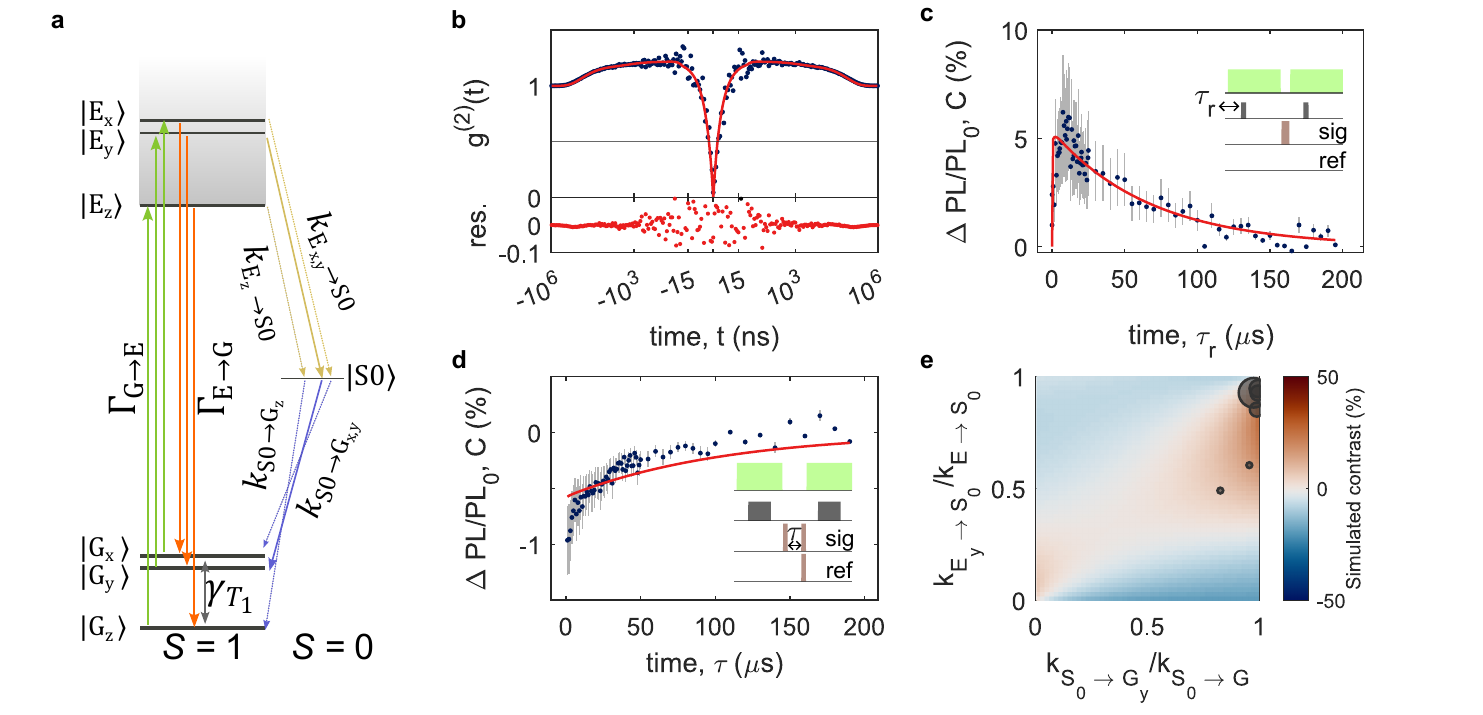}
  \caption{\textbf{Optical and spin dynamics of carbon-related hBN defects}. \textbf{a}  Model used to fit the results of experiments in \textbf{b}-\textbf{d}. The model includes a spin-1 ground and optically excited states and a singlet metastable state. We assume that optical excitation and radiative recombination processes are spin conserving at zero magnetic field. In panels \textbf{b-d}, red curves correspond to a global fit of the model to the results of all three experiments. \textbf{b} (top) Background-corrected second-order autocorrelation ($g^{(2)}(t)$) (blue circles). (bottom) Residuals of the fit of the model to the data. \textbf{c} Spin-dependent optical initialisation. The inset presents the pulse protocol with optical (green blocks) and microwave drive pulses (red blocks) and readout time (gray block). Blue circles are the mean value of the contrast measured for various delay times $\tau_\text{r}$. Error bars indicate standard error of mean. \textbf{d} Modified spin-relaxation experiment. The signal experiment probes the PL when we apply two microwave $\pi$ pulses, each before and after delay time $\tau$ between the two optical pulses. The reference experiment probes the PL when a single microwave $\pi$ pulse is applied at the end of $\tau$. Blue circles are the mean value of the measured contrast, with one standard deviation indicated by error bars. \textbf{e} Simulated ODMR contrast as a function spin-selective direct ($k_{E \rightarrow S0}$) and reverse ($k_{S0 \rightarrow G}$) intersystem crossing rates. Black circles represent data measured for different defects, with the position indicating relative rates extracted from PECS and pulsed ODMR experiments, and size corresponding to measured cwODMR contrast. The largest circle corresponds to 30~\% contrast. 
  \label{Fig::fig2}}
\end{figure*}

For optically active spin defects, the ODMR contrast is dependent on the degree of spin initialisation arising from the optical cycle. 
Across the defects we study, we observe the magnitude of the saturated zero-field ODMR contrast across the three spin resonances typically follows: \textit{C}(\textit{f}\textsubscript{A}) = \textit{C}(\textit{f}\textsubscript{B}) \textgreater \textit{C}(\textit{f}\textsubscript{C}) with defect-to-defect variation in overall magnitude \cite{Stern2024} (see Fig.~S2). 
This observation is in line with an optical defect type that shows variable ISC rates, consistent with the variation we see in bunching timescales in second-order autocorrelation ($g^{(2)}(t)$) experiments.
To understand the remarkably high ODMR contrast and its retention with off-axis field for the hBN defects, we investigate the optical rates of the system by setting up a series of rate equations describing the transfer of population between the electronic states for the model shown in Fig.~\ref{Fig::fig2}a in the absence of a magnetic field.
The non-equal ODMR contrast of \textit{f}\textsubscript{B} and \textit{f}\textsubscript{C} indicates asymmetry of the intersystem crossing
rates into $\ket{\text{G\textsubscript{x}}}$ and $\ket{\text{G\textsubscript{y}}}$ (from $\ket{\text{E\textsubscript{x}}}$ and $\ket{\text{E\textsubscript{y}}}$) eigenstates at zero-field, as is observed for other low symmetry $S$=1 systems \cite{kohler,Wrachtrup1993,Mena2024}. 
In our kinetic model we hold $k_{\text{E\textsubscript{x}}\rightarrow S0} = k_{\text{E\textsubscript{z}}\rightarrow S0} \neq k_{\text{E\textsubscript{y}}\rightarrow S0}$ and $k_{S0\rightarrow \text{G\textsubscript{x}}} = k_{S0\rightarrow \text{G\textsubscript{z}}} \neq k_{S0\rightarrow \text{G\textsubscript{y}}}$ in order to restrict the number of fitting parameters, but note that some defects are best described by $k_{\text{E\textsubscript{x}}\rightarrow S0} \neq k_{\text{E\textsubscript{z}}\rightarrow S0} \neq k_{\text{E\textsubscript{y}}\rightarrow S0}$ ($k_{S0\rightarrow \text{G\textsubscript{x}}} \neq k_{S0\rightarrow \text{G\textsubscript{z}}} \neq k_{S0\rightarrow \text{G\textsubscript{y}}}$).

We determine the optical rates for a second single defect at zero field via a global fit to the combined results of the second-order autocorrelation ($g^{(2)}(t)$, Fig.~\ref{Fig::fig2}b) and pulsed ODMR measurements (Fig.~\ref{Fig::fig2}c,d).
The pulsed ODMR sequences are illustrated in the insets of Figs.~\ref{Fig::fig2}c,d, where the microwave pulses are $\pi$ pulses calibrated via Rabi experiments on resonance with \textit{f}\textsubscript{B} (see Fig.~S3). 
Figure~\ref{Fig::fig2}b shows the background-corrected $g^{(2)}(t)$ measurement for this defect (see Supplementary Materials, Sec.~III for details on the background correction procedure). 
The horizontal (time) axis is presented in linear scale between -30 and 30~ns, where we can see the characteristic antibunching dip at $t=0$. 
For $|t| > 30$~ns, we present the time axis in log scale. 
The hBN defects show significant bunching behavior, indicative of the presence of a long-lived metastable state, which only subsides after $\sim$100~$\mu$s. 
Similar trends have been reported for various types of hBN emitters \cite{Stern2022,Patel2022,Patel2023,Exarhos2019,Hamidreza2024,Zhong2024}.\\


Figure~\ref{Fig::fig2}c presents the dynamics of the spin-dependent optical initialisation \cite{Patel2023} of the same hBN defect presented in Fig.~\ref{Fig::fig2}b, while Fig.~\ref{Fig::fig2}d presents its spin-relaxation dynamics. 
For Figs.~\ref{Fig::fig2}b,~\ref{Fig::fig2}c and~\ref{Fig::fig2}d the red curves are the result of a global fit of the optical model (Fig.~\ref{Fig::fig2}a) to the experimental data and Tab.~\ref{Tab::tab1} presents the corresponding rates extracted from this fit (see Supplementary Materials, Sec.~IV, for details on model and fitting procedure). 
We note in our analysis we also considered a model with singlet ground and optically excited states and a triplet metastable state, but it fails to capture the observed behaviour (see Supplementary Materials, Sec.~V). 
For this defect, the global fit reveals comparable magnitudes for the radiative ($\Gamma_{\text{E} \rightarrow \text{G}}=163$~MHz) and non-radiative ($ k_{\text{E} \rightarrow S0} = \sum\limits_{i = x,y,z}{k_{\text{E\textsubscript{i}} \rightarrow S0}} = 200$~MHz) decay rates from the optically excited state, and strongly spin-selective direct and reverse intersystem crossing ($k_{\text{E\textsubscript{y}} \rightarrow S0}/k_{\text{E} \rightarrow S0}$ = 0.9462 and $k_{S0 \rightarrow \text{G\textsubscript{y}}}/k_{S0 \rightarrow \text{G}}$ = 0.9941). \\
 

We repeat this procedure for 5 defects with the same zero-field splitting resonance and find that, while the magnitude of the radiative and intersystem crossing rates are broadly similar across defects, there is significant variation in the ratio of spin-dependent intersystem crossing rates ($k_{\text{E\textsubscript{y}} \rightarrow S0}/k_{\text{E} \rightarrow S0}$ = 0.49-0.95, $k_{S0 \rightarrow \text{G\textsubscript{y}}}/k_{S0 \rightarrow \text{G}}$ = 0.82-0.99). 
This explains the variation (from \textless 1\% to 95\%) in the magnitude of the saturated ODMR contrast per defect \cite{Stern2024} (see Supplementary Materials, Sec.~VI for extended data from which individual rates are extracted). 
Figure~\ref{Fig::fig2}e shows the interdependence of the cwODMR contrast on the spin-selectivity of the direct ($k_{\text{E} \rightarrow S0}$, vertical axis) and reverse ($k_{S0 \rightarrow \text{G}}$, horizontal axis) intersystem crossing rates.
The 2D map presents the simulated cwODMR contrast of \textit{f}\textsubscript{B}, where the rates indicated in the axes are varied while all remaining rates are kept constant at the values presented in Tab.~\ref{Tab::tab1}. 
The colour represents the amplitude of cwODMR contrast predicted by the model, with red (blue) regions indicating positive (negative) contrast. 
The black circles show the experimental cwODMR contrast for each defect we measured (where the size represents the magnitude of cwODMR contrast, see Supplementary Material, Sec.~VII for the raw spectra), positioned on the map as a function of the determined rates for each defect. 
The rates extracted using the procedure outlined above for the hBN defects cluster in the top right of the 2D plot, showing that these defects are characterised by strong spin-selectivity in both direct and reverse intersystem crossing processes.  
As a result, spin mixing requires a larger applied magnetic field in order to disrupt the optical spin initialisation mechanism \cite{Tetienne2012}, giving rise to large magnetic field dynamic range for the hBN sensor.\\

\textbf{High sensitivity vectorial magnetic-field sensing enabled by the low symmetry spin}

To model the sensitivity of the hBN spins as a function of orientation and strength of applied magnetic field, we include the effect of magnetic field in the model by introducing the Zeeman term to the spin Hamiltonian ($H_\text{ZE}$). We determine the magnetic-field dependent intersystem crossing rates from a statistical average of the zero-field rates, such that $k_{ij}(\mathbf{B}) = \sum_{p,q} |a_{ip}|^2 |a_{jq}|^2 k_{pq}^0$, similar to the approach taken by Epstein \textit{et al.} and Tetienne \textit{et al.} for the NV centre in diamond \cite{Epstein2005anisotropic,Tetienne2012}. Here, $k_{pq}^0$ are the zero-field direct and reverse spin-dependent intersystem crossing rates; the coefficients $a_{ip}$ can be obtained by comparing the zero-field eigenstates ($\ket{p(0)}$) to the eigenstates of the Hamiltonian at a field ($\ket{i(\mathbf{B})}$), such that $\ket{i(\mathbf{B})} = \sum_{p} a_{ip} \ket{p(0)}$. In the absence of spectroscopic information about the excited-state zero-field splitting configuration, we assume equal zero-field splitting parameters in ground and optically excited states. This assumption has no significant implication on the findings of this work (see Supplementary Material, Sec.~VIII). \\

Figure~\ref{Fig::fig3}a presents the evolution of the ground-state spin eigenstates for the hBN defect system under applied magnetic field in the $x, y, z$ direction (top to bottom panels).
The purple circles represent the simulated optically initialised population, calculated based on the model above and using the representative rates of Tab.~\ref{Tab::tab1}.
In the zero-field limit, the system is initialised into the $\ket{\text{G\textsubscript{y}}}$ state, a direct consequence of the low symmetry observed in this system, giving rise to strong \textit{f}\textsubscript{A} and \textit{f}\textsubscript{B} and weak \textit{f}\textsubscript{C} (Fig.~\ref{Fig::fig1}). 
Magnetic field applied along the defect \textit{y} axis (middle panel) mixes $\ket{\text{G\textsubscript{x}}}$ and $\ket{\text{G\textsubscript{z}}}$, preserving the zero-field character of $\ket{\text{G\textsubscript{y}}}$, thus retaining the zero-field spin initialisation and ODMR contrast. 
Conversely, applied field along $x$ ($z$) mixes $\ket{\text{G\textsubscript{y}}}$ and $\ket{\text{G\textsubscript{z}}}$ ($\ket{\text{G\textsubscript{x}}}$), redistributing the zero-field initialised population and modifying the saturated cwODMR contrast of each resonance with respect to their zero-field values. 
Importantly, this model indicates that a field can be applied along any of the three defect axes and cwODMR contrast is retained. \\

We confirm the predictions of this model by investigating the magnetic-field dependence of the cwODMR for the same defect shown in Figure~\ref{Fig::fig1}. 
Figures~\ref{Fig::fig3}b,c show, respectively, the dependence of cwODMR central frequencies (top panel) and normalized cwODMR contrast of \textit{f}\textsubscript{A}-\textit{f}\textsubscript{C} (bottom panels), on the orientation of 51-mT magnetic field in the $yz$ (b) and $xy$ (c) planes, and a comparison to the predictions of the model (solid curves). 
We normalize the cwODMR contrast by the zero-field cwODMR contrast of \textit{f}\textsubscript{B}. 
This allows us to compare the magnetic-field dependence of contrast observed experimentally and calculated theoretically for defects with different values of zero-field contrast. 
The data shows that applied field along the defect $y$ axis (indicated by 90 degrees in Fig.~\ref{Fig::fig3}b and c) preserves the zero-field contrast distribution.
The cwODMR contrast of \textit{f}\textsubscript{A} (\textit{f}\textsubscript{B}) is completely (partially) suppressed as the magnetic field is rotated towards the $z$ axis, while the cwODMR contrast of \textit{f}\textsubscript{C} increases (Fig.~\ref{Fig::fig3}b). 
We note that the sharp dip in contrast of \textit{f}\textsubscript{A} and \textit{f}\textsubscript{B} when the field is applied directly along the $y$ axis is reproducible, but we have not identified its origin. 
Rotation of the applied field in the $xy$ plane away from the $y$ axis leads to a slower suppression of the cwODMR contrast of both \textit{f}\textsubscript{A} and \textit{f}\textsubscript{B}, with a correspondingly slower increase of the cwODMR contrast of \textit{f}\textsubscript{C}. 
Figure~\ref{Fig::fig3}d presents the cwODMR spectra as a function of $B_y$ amplitude up to 140~mT, showing that for this class of defects contrast is preserved for an applied field along the defect's $y$ axis. These results elucidate that a low symmetry $S$=1 system has three inequivalent axes \cite{Richert2017}. This is advantageous for magnetometry, as the optical readout of the spin resonances is preserved for arbitrarily oriented magnetic fields. \\


We use our understanding of the field-dependent photophysics of the hBN defects to determine the expected DC magnetic-field sensitivity as a function of orientation and strength of applied magnetic field, \textit{i.e.} the operating range of this sensor. 
DC sensitivity is given by the relationship $\eta_{DC} = \alpha \frac{1}{\partial \nu_i/\partial B} \frac{\Delta \nu}{C \sqrt{\text{PL}_0}}$, where $\alpha$ is a prefactor associated with the cwODMR lineshape ($\alpha = \sqrt{e/(8 \log{2})}$ for a Gaussian lineshape), $\partial \nu_i/\partial B$ is the resonance frequency dependence on magnetic-field amplitude, $\Delta \nu$ is the cwODMR resonance full width at half maximum, $C$ is the contrast and $\text{PL}_0$ is the brightness of the defect in the absence of microwaves corrected by a factor $0.1$ to account for collection losses \cite{Oshnik2022,Barry2020}. We use $\Delta \nu \approx 30$~MHz, extracted from the cwODMR spectra of Fig.~\ref{Fig::fig1}e, and $\partial \nu_i/\partial B$, $C$ and $\text{PL}_0$ predicted by the Hamiltonian in Eq.~\ref{Eq::HamiltonianI} and the model in Fig.~\ref{Fig::fig2}a, with rates from Tab.~\ref{Tab::tab1}. \\

Nanoscale magnetometry typically requires the application of bias magnetic field to distinguish between positive and negative orientations of target field \cite{Tan2023,Balasubramanian2008,Rondin2012,Degen2008}. 
As we have explained above, for the NV centre in diamond a bias field must be carefully aligned along the $z$ axis to retain ODMR contrast. 
We show here how the anisotropic cwODMR response of the hBN defects relaxes these stringent requirements on the bias-field orientation. 
Figure~\ref{Fig::fig4}a presents the calculated cwODMR sensitivity ($\eta_{DC}$) of each resonance \textit{f}\textsubscript{A}-\textit{f}\textsubscript{C} as a function of bias-field orientation. 
We plot the calculated sensitivity as colour on a sphere, where the sphere indicates the direction of the 50-mT bias field orientation. 
$B_{0,x}$, $B_{0,y}$ and $B_{0,z}$ axes indicate bias field aligned to the $x,y,z$ axes of the defect, respectively. 
In this way, coloured regions indicate configurations with high sensitivity, whereas gray regions indicate configurations with vanishing sensitivity. 
In agreement with the results presented in Fig.~\ref{Fig::fig3}, we predict that this system shows at least one of the three possible spin resonances at arbitrary 50-mT bias-field orientation, enabling sub-$10~\mu$T/$\sqrt{\text{Hz}}$ sensitivity independent of orientation of bias field, and down to 500~nT/$\sqrt{\text{Hz}}$ for optimal bias-field alignment. 
For comparison, Fig.~\ref{Fig::fig4}b presents the simulated sensitivity of the NV centre in diamond, calculated using optical rates from Ref.~\cite{Tetienne2012}. The coloured region around the $B_{0,z}$ axis reflects the fact that, as expected, the NV-centre sensitivity lies predominantly along its high-symmetry axis. \\

Full vectorial sensing relies on unambiguous determination of three linearly independent target magnetic field components. 
For NV magnetometry, the target field vector is reconstructed from its projection along the $z$ axis of four site inequivalent NVs \cite{Tetienne2017,Broadway2020,Chen2020}. 
In Fig.~\ref{Fig::fig4}c, we combine the results from (a-c) to present the calculated vectorial target-field sensitivity for the hBN defects for arbitrary bias-field orientation and assess the possibility of this system to provide vectorial information using only one defect. 
As indicated in the inset, the direction of the coloured arrows indicate the direction of maximum sensitivity of each cwODMR resonance, given by the gradient of the resonance frequency ($\nabla f_\text{i}(B_x,B_y,B_z)$). The size of the arrows corresponds to inverse absolute sensitivity of each cwODMR resonance. 
As above, the sphere represents the orientation of the 50-mT bias field. 
We find that via two measurements, utilising two bias-field orientations (shown via the black arrows, labelled (1) and (2)) the full target-field vector can be reconstructed. 
This is because a bias field along (1) allows determination of $y$ and $z$ components of the target field and a bias field along (2) allows determination of a field along $x$. In this way, this system enables vectorial mapping of target magnetic field with two independent measurements at different bias-field configurations, using one stationary defect. 
We note that this approach differs from existing protocols for vectorial magnetic field sensing using defects in silicon carbide and diamond as it does not require physical rotation of the quantum sensor itself \cite{Lee2015,Niethammer2016,stefan2021}. \\

\textbf{Outlook}

Our results present a new candidate for nanoscale quantum sensing with intrinsic characteristics of low symmetry and high spin contrast and brightness that makes it competitive with state-of-the-art defect systems for DC magnetometry. 
In addition to the large dynamic range,  $500~$nT/$\sqrt{\text{Hz}}$ DC sensitivity and vectorial sensing capabilities described above, the 2D system has a natural advantage with regards to sample-sensor proximity. The defects can be grown into multilayers only a few-nm thick and the 2D host enables simple integration into 2D heterostructures, as well as established tip-based sensing approaches, which opens routes to achieving well below 20-nm spatial resolution with this platform. In addition, the hBN host is inert in biological media \cite{kavvcivc2024}, inexpensive to produce and can be grown to scale. \\

We postulate that the 2D host material may offer future advantages for optimal tuning the dynamics of defects, in-situ. Interestingly, our kinetic modelling and ODMR results across defects of this type show the magnitude and ratio of optical rates depend on the defect in question, despite a well-defined ground state (see Supplementary Material, Sec.~IV for statistics on radiative and non-radiative decay rates across defects). In particular, we find that persistence of contrast over a broad magnetic field regime arises from the strong spin selectivity of both the direct and reverse intersystem crossing rates of the system. We predict the excited-state dynamics in this system are highly sensitive to local strain and electric field, due to being embedded in a 2D material that can be highly strained. Despite being a source of inhomogeneous behaviour, this feature could be harnessed as a pathway to control and enhance contrast of individual emitters via strain or electric field tuning, for example. Finally, these results pertain mainly to the performance of this system for DC sensing. However, the spin properties of this hBN defect, including microsecond spin coherence at room temperature that can be accessed via dynamical decoupling protocols \cite{Stern2024}, open routes for exploring the system for AC sensing in the future. Combined, these results demonstrate the potential of hBN defects for quantum sensing. \\

\textbf{Experimental methods}

\textit{Materials}

Multilayer hBN was grown by metal-organic vapour phase epitaxy on sapphire substrates and subsequently transferred to Si/SiO\textsubscript{2} using water-assisted self-delamination. Details of the growth process have been provided extensively elsewhere \cite{Chugh2018}. Triethyl boron and ammonia were used as boron and nitrogen precursors during MOVPE, where the flow rate of triethyl boron has been shown to be correlated to the incorporation of carbon and the observation of visible emitters in the resulting hBN \cite{Mendelson2021}.  \\

\textit{Confocal photoluminescence microscopy}

Photoluminescence measurements are conducted at room temperature on two home-built free-space confocal microscopy setups. For both cases, experimental hardware is connected to a data acquisition card (National Instruments, PCIe6323), controlled via open-source Python suite Qudi \cite{binder_qudi_2017}.
We use a 532-nm continuous wave laser (Ventus 532, Laser Quantum), split into several excitation lines using a beamsplitter, where each path is directed to a different setup. In each line, the laser passes through an acousto-optic modulator (AA Optoelectronics). The first-order diffracted beam is fibre-coupled into the relevant optical setup, enabling intensity modulation of the laser light. The laser is filtered using a 532-nm laser line filter (Thorlabs, FL532-3) after out-coupling into the free space microscopy setup, to eliminate any fibre-related emission. The beam passes through a beamsplitter (Thorlabs, 90:10 R:T), with the reflected beam providing sample excitation and the intensity of the intensity of the transmitted beam is monitored using a photodiode, completing a feedback loop that allows laser-power control in conjunction with the AOM. Confocal scanning of the sample is enabled by a scanning mirror (Physik Instrumente, S-334.2SL), and a 100x 0.9 NA air objective (Nikon Instruments). The emitted light travels through the setup in reverse path to the excitation, and is coupled into a single-mode fibre (Thorlabs, SM600) after passing through two 550-nm long-pass filters (FEL550, Thorlabs) to remove the laser light. The collection fibre is coupled into either a single-photon avalanche photodiode (SPCM-AQRH-14-FC, Excelitas Technologies) or into a charge-coupled-device-coupled spectrometer (Acton Spectrograph, Princeton Instruments). \\

\textit{Optically detected magnetic resonance}

ODMR measurements are conducted using confocal microscopy as described above. Microwaves are produced using an RF signal generator (Stanford Research Systems DC to 4.05 GHz Signal Generator or Marconi Instruments 10 kHz to 2.4 GHz Signal Generator), amplified (ZHL-42 W+, 0-4.2 GHz, 30-35 dB, MiniCircuits), and delivered to the sample by a homemade copper loop antenna placed between the objective lens and the hBN multilayer, with the confocal spot approximately aligned to the centre of the loop to allow optical access.

During cwODMR, the 532-nm excitation is continuously applied, whilst the intensity of the microwaves is modulated using a 140-Hz square wave, and the microwave frequency is modulated at 7~Hz. The PL count rate is monitored as a function of microwave amplitude and frequency to calculate ODMR contrast, representing the fractional change in PL counts upon the application of the spin-flipping microwave drive ($C = (\text{PL}_{\text{sig}}-\text{PL}_{\text{ref}})/\text{PL}_{\text{ref}}$). The contrast is calculated per datapoint, and finally averaged for each microwave frequency position across all measurement sweeps.

For pulsed ODMR measurements, we use a pulse streamer (Swabian 8/2) to control a series of switches (MiniCircuits ZYSWA-2-50-DR+). The switches allow the generation of square-wave laser and microwave pulses by modulating the AOM trigger level and the connection between the RF source output and the antenna, respectively. A switch also controls the signal readout duration. The pulse sequences used during measurements are described in detail in the text, the Supplementary information, and previous work \cite{Stern2024}.

Angle-resolved magnetic field is applied using a dual-axis homebuilt mount, consisting of two rotation stages (Thorlabs) and a permanent magnet on a custom mount. Zero-field measurements are conducted without shielding, in the Earth's magnetic field. \\

\textit{Hanbury-Brown Twiss Interferometry}

Intensity autocorrelation measurements are conducted using Hanbury-Brown-Twiss interferometry. The fluorescence collection fibre is connected to a 50:50 fibre beamsplitter, with each end coupled into a single-photon avalanche photodiode. The PL counts at each photodiode are monitored by a time-to-digital converter (quTAU, qutools) with 81-ps resolution. \\

\textbf{Acknowledgements}

We thank Toby Mitchell for technical assistance, and Viktor Iv\'ady, Sam Bayliss, and Qiushi Gu for helpful discussions.
We acknowledge support from the ERC Advanced Grant PEDESTAL (884745) and the Office of Naval Research Global (N62909-22-1-2028). C.M.G. acknowledges support from the Netherlands Organisation for Scientific Research (NWO 019.221EN.004, Rubicon 2022-1 Science). S.E.B., S.A.F. and O.F.J.P. acknowledge funding from the EPSRC Centre for Doctoral Training in Nanoscience and Nanotechnology (NanoDTC, grant no. EP/S022953/1). C.L.C. acknowledges funding from the EPSRC Doctoral Training Programme and Vice-Chancellor Award. D.L. acknowledges scholarships from the Skye Foundation and the Cambridge Trust. C.L., I.A., and H.H.T. acknowledge funding from the Australian Research Council, through grants CE200100010 and FT220100053. H.L.S. acknowledges a Royal Society fellowship.

\FloatBarrier

\begin{table}
    \centering
    \caption{\textbf{Model parameters.} Summary of key parameters obtained from fitting the data in Figs.~\ref{Fig::fig2}b-d to the model.}
    \label{Tab::tab1}
    \begin{tabular}{|c|c|c|c|c|c|c|c|c|c|c|}
        \hline
         Defect & $\Gamma_{G\rightarrow E}$ & $\Gamma_{E\rightarrow G}$ & $k_{\text{E\textsubscript{x}} \rightarrow S0}$ & $k_{\text{E\textsubscript{y}} \rightarrow S0}$ & $k_{\text{E\textsubscript{z}} \rightarrow S0}$ & $k_{S0 \rightarrow \text{G\textsubscript{x}}}$ & $k_{S0 \rightarrow \text{G\textsubscript{y}}}$ &$k_{S0 \rightarrow \text{G\textsubscript{z}}}$ & $\gamma_{T_1}$ & Contrast\\
         Unit & kHz/$\mu$W & MHz & MHz & MHz & MHz & kHz & kHz & kHz  & kHz & \% \\
         \hline \hline
         J19 &0.92 &	163.4&	5.4&	190&	5.4&	2&	675&	2&	3.2 & 22\\
         \hline
    \end{tabular}
\end{table}

\begin{figure*}[t]
  \includegraphics[width=\textwidth]{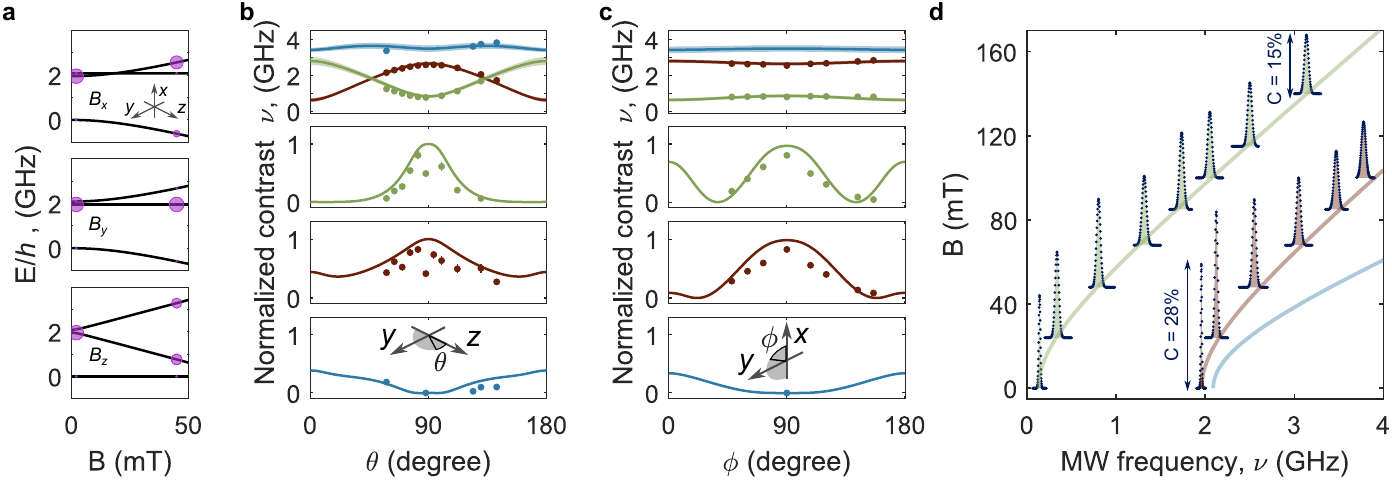}
  \caption{\textbf{Magnetic-field orientation and amplitude dependence of cwODMR}. \textbf{a} Evolution of spin eigenstates of the Hamiltonian Eq.~\ref{Eq::HamiltonianI} with applied magnetic field along the \textit{x}, \textit{y}, \textit{z} axes of the defect, from top to bottom. Magnetic field mixes the zero-field spin eigenstates, modifying the optically initialised population. Calculated amplitudes of the optically initialised population of each spin sublevel are indicated by the size of the purple circles.
  \textbf{b,c} Angular magnetic-field dependence of cwODMR frequency (top panels, see SI for raw spectra) and contrast of resonances \textit{f}\textsubscript{A} to \textit{f}\textsubscript{C}, normalized by the zero-field cwODMR contrast of the \textit{f}\textsubscript{B} resonance. Data are presented as circles, with colour coding according to the inset of Fig.~\ref{Fig::fig1}d, and curves indicate the cwODMR contrast simulated using the model of Fig.~\ref{Fig::fig2} and fit parameters presented in Tab.~\ref{Tab::tab1}. Insets indicate the direction of rotation (in the \textit{yz} plane for \textbf{b}, and in the \textit{xy} plane for \textbf{c}). In \textsubscript{b}, $\theta$ varies with fixed $\phi=90\degree$; in \textbf{c}, $\phi$ varies with fixed $\theta=85\degree$. The field amplitude and rotation range were limited by the experimental geometry. \textbf{d} Persistence of saturated cwODMR contrast for a field applied along the defect $y$ direction ($\phi=90\degree$,$\theta=85\degree$), shown up to 140~mT. The solid curves represent the transition frequencies of \textit{f}\textsubscript{A} (green), \textit{f}\textsubscript{B} (red) and \textit{f}\textsubscript{C} (blue) resonances as a function of $B_y$ amplitude. The cwODMR spectra are represented by blue circles, measured up to 140~mT. \label{Fig::fig3}}
\end{figure*}

\begin{figure*}[t]
  \includegraphics[width=\textwidth]{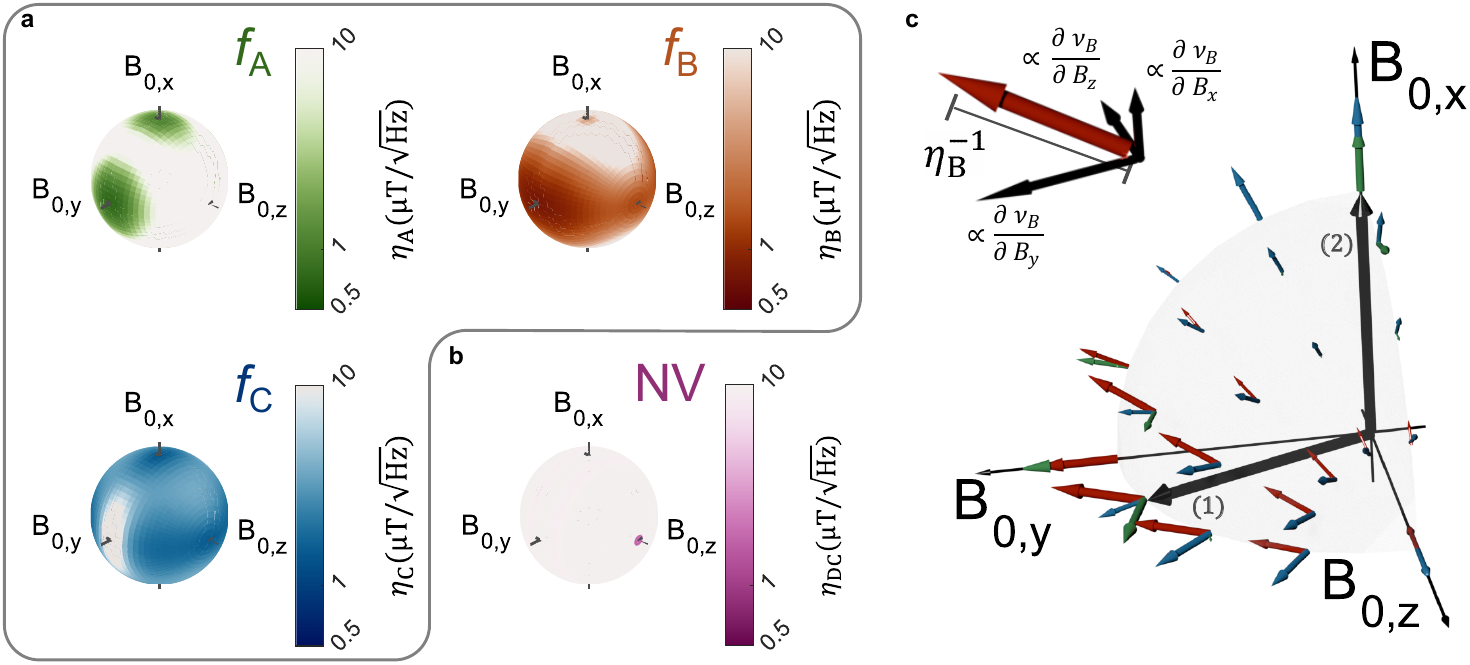}
  \caption{\textbf{Magnetic-field sensitivity range}.
  \textbf{a} Simulated magnetic-field sensitivity of each cwODMR resonance of the carbon-related defect in hBN as a function of the orientation of 50-mT bias magnetic field. The spheres indicate the magnetic-field orientation, with $B_{0,x}$, $B_{0,y}$, $B_{0,z}$ respectively aligned parallel to the $x, y, z$ axis of the defect, whereas the colour scale indicates the target-field sensitivity. \textbf{b} Simulated magnetic-field sensitivity of the NV centre in diamond as a function of the orientation of 50-mT bias magnetic field. \textbf{c} Direction of target field providing optimal sensitivity for each cwODMR resonance as a function of bias-field orientation. As indicated in the top left inset, the coloured arrows' size corresponds to the inverse sensitivity of resonances \textit{f}\textsubscript{A-C}, with colour coding according to Fig.~\ref{Fig::fig1}d. The orientation of each coloured arrow is given by the gradient of the resonance frequency of resonances \textit{f}\textsubscript{A-C} for a given bias-field configuration, indicating the direction of optimal sensitivity. 
  \label{Fig::fig4}}
\end{figure*}

\clearpage

\bibliography{MyLibrary}

\end{document}


\renewcommand{\thefigure}{S\arabic{figure}}
\renewcommand{\theequation}{S\arabic{equation}}




\title{Supplementary Information for: A single spin in hexagonal boron nitride for vectorial quantum magnetometry}

\author{Carmem~M.~Gilardoni}
\altaffiliation{These authors contributed equally to this work} 
\affiliation{Cavendish Laboratory, JJ Thomson Avenue, University of Cambridge, Cambridge CB3 0HE, UK}
\author{Simone~Eizagirre~Barker}
\altaffiliation{These authors contributed equally to this work} 
\affiliation{Cavendish Laboratory, JJ Thomson Avenue, University of Cambridge, Cambridge CB3 0HE, UK}
\author{Catherine~L.~Curtin}
\affiliation{Cavendish Laboratory, JJ Thomson Avenue, University of Cambridge, Cambridge CB3 0HE, UK}
\author{Stephanie~A.~Fraser}
\affiliation{Cavendish Laboratory, JJ Thomson Avenue, University of Cambridge, Cambridge CB3 0HE, UK}
\author{Oliver.~F.J.~Powell}
\affiliation{Cavendish Laboratory, JJ Thomson Avenue, University of Cambridge, Cambridge CB3 0HE, UK}
\affiliation{Hitachi Cambridge Laboratory, Hitachi Europe Ltd., JJ Thomson Avenue, Cambridge CB3 0HE, UK}
\author{Dillon~K.~Lewis}
\affiliation{Cavendish Laboratory, JJ Thomson Avenue, University of Cambridge, Cambridge CB3 0HE, UK}
\author{Xiaoxi~Deng}
\affiliation{Cavendish Laboratory, JJ Thomson Avenue, University of Cambridge, Cambridge CB3 0HE, UK}
\author{Andrew~J.~Ramsay}
\affiliation{Hitachi Cambridge Laboratory, Hitachi Europe Ltd., JJ Thomson Avenue, Cambridge CB3 0HE, UK}
\author{Chi~Li}
\affiliation{ARC Centre of Excellence for Transformative Meta-Optical Systems, Faculty of Science, University of Technology Sydney, Ultimo, New South Wales, Australia}
\affiliation{School of Mathematical and Physical Sciences, Faculty of Science, University of Technology Sydney, Ultimo, New South Wales, Australia}
\author{Igor~Aharonovich}
\affiliation{ARC Centre of Excellence for Transformative Meta-Optical Systems, Faculty of Science, University of Technology Sydney, Ultimo, New South Wales, Australia}
\affiliation{School of Mathematical and Physical Sciences, Faculty of Science, University of Technology Sydney, Ultimo, New South Wales, Australia}
\author{Hark~Hoe~Tan}
\affiliation{ARC Centre of Excellence for Transformative Meta-Optical Systems, Department of Electronic Materials Engineering, Research School of Physics, The Australian National University, Canberra, ACT 2600, Australia}
\author{Mete~Atat\"ure}
\affiliation{Cavendish Laboratory, JJ Thomson Avenue, University of Cambridge, Cambridge CB3 0HE, UK}
\author{Hannah~L.~Stern}
\affiliation{Photon Science Institute, Department of Physics and Department of Chemistry, The University of Manchester, Manchester, M13 9PL, UK}

\date{Version of \today}

\let\thefootnote\relax\footnotetext{Correspondence to: CMG (cm2207@cam.ac.uk) or HLS (hannah.stern@manchester.ac.uk)}

\maketitle

\tableofcontents

\newpage






\section{Variations in cwODMR contrast between defects}


\begin{figure}[h!]
    \centering
    \includegraphics[width=0.8\linewidth]{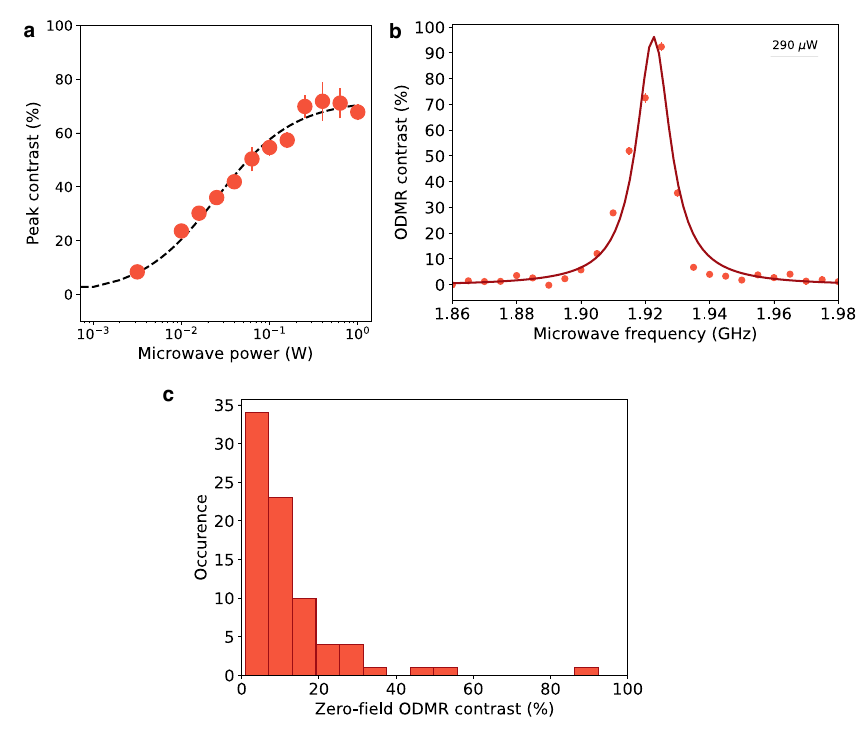}
    \caption{\textbf{Variations in zero-field contrast across hBN defects.} (a) cwODMR contrast \textit{f}\textsubscript{B} resonance as a function of microwave power showing clear saturation behaviour, for a defect illuminated with 150$\mu$W optical power. (b) cwODMR spectrum of the  \textit{f}\textsubscript{B} resonance same defect, taken highest maximum microwave (1.5~mW) and optical powers accessible (290~$\mu$W). The contrast reaches $>$90\%. (c) Histogram of the maximum cwODMR contrast observed for 79 defects in the absence of a magnetic field. These measurements were taken at 1.5~mW microwave power.}
    \label{SIFig::contrast_stats}
\end{figure}

The saturated cwODMR contrast varies across defects, ranging from 1-2\% to over 90\%. In Figure~\ref{SIFig::contrast_stats}a, the peak cwODMR contrast is presented as a function of microwave power, exhibiting clear saturation behaviour. In Figure~\ref{SIFig::contrast_stats}b, the cwODMR contrast of the \textit{f}\textsubscript{B} resonance of this defect reaches $>$90\%.
The distribution of saturated contrast across defects is represented by the histogram in Figure~\ref{SIFig::contrast_stats}c. These values correspond to peak cwODMR contrast of the \textit{f}\textsubscript{B} resonance, measured under microwave saturation conditions. However, a full laser power-dependence of the contrast was not conducted for all defects presented, such that the value of contrast quoted here presents a lower bound.

The relative distribution of contrast between the three possible cwODMR resonances \textit{f}\textsubscript{A}, \textit{f}\textsubscript{B}, and \textit{f}\textsubscript{C} is dependent on the defect. For all defects, we observe the highest magnitude of contrast into the \textit{f}\textsubscript{B} resonance. For many, we see comparable magnitude between \textit{f}\textsubscript{A} and \textit{f}\textsubscript{B} (as shown below), while for others contrast is more evenly distributed between \textit{f}\textsubscript{A} and \textit{f}\textsubscript{C}.

\begin{figure}
    \centering
    \includegraphics[width=0.75\linewidth]{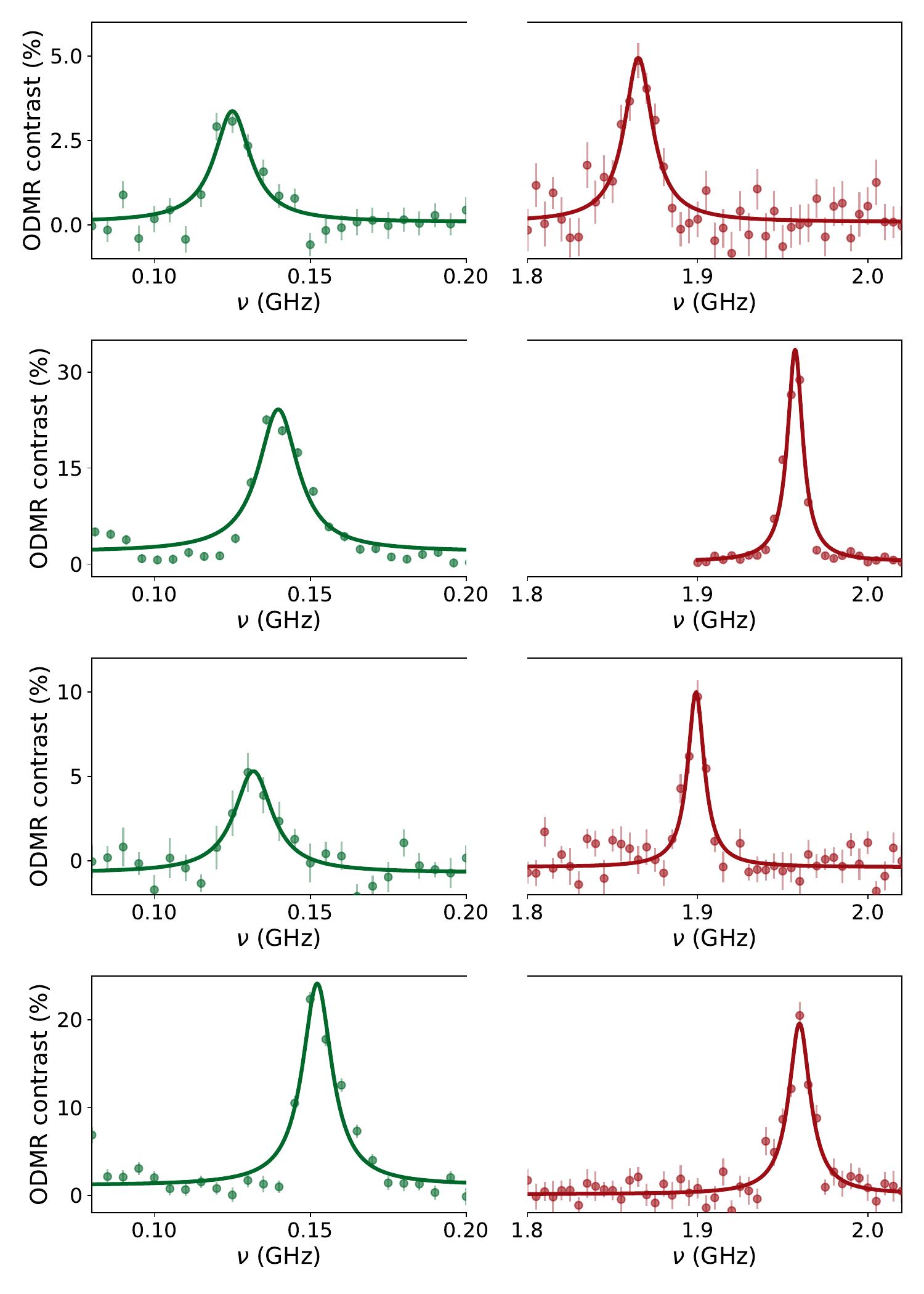}
    \caption{\textbf{Saturated cwODMR spectra for four defects.} For these defects, the contrast of \textit{f}\textsubscript{A} is comparable to that of \textit{f}\textsubscript{B}, while the contrast of \textit{f}\textsubscript{C} is lower than the signal-to-noise ratio of the measurement.}
    \label{SIFig::contrast_distribution}
\end{figure}

\FloatBarrier
\newpage
\section{Rabi measurement}

\begin{figure}[h]
    \centering
    \includegraphics[width=0.6\linewidth]{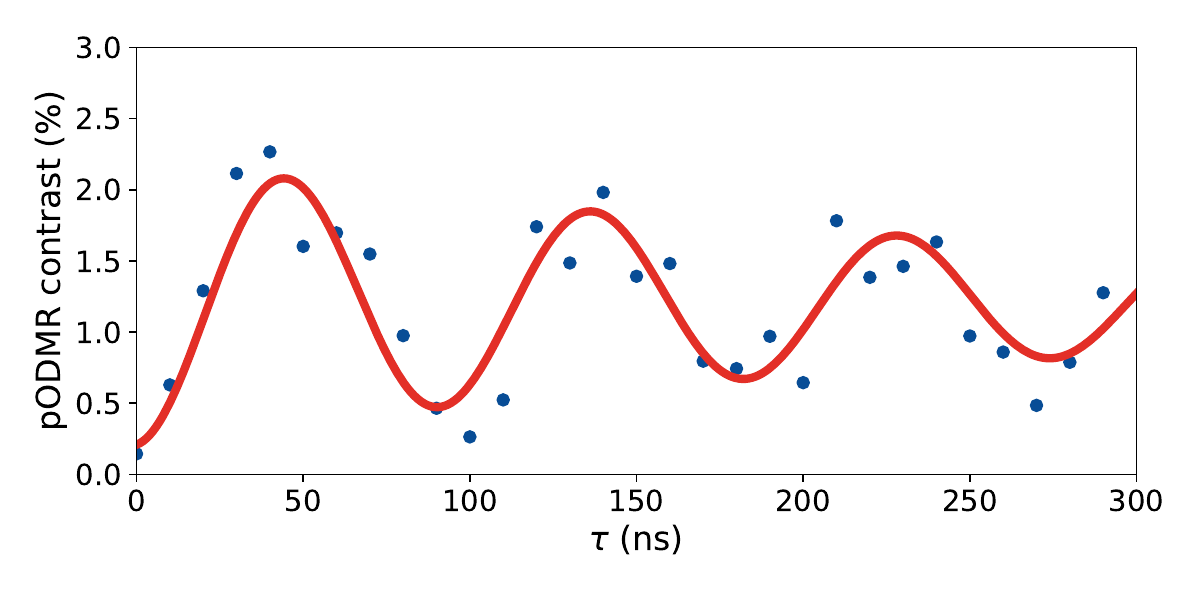}
    \caption{Rabi measurement for the defect presented in Main Text, Fig.~2, used to calibrate the duration of $\pi$ pulses in subsequent experiments.}
    \label{fig:T1pipulseanalysis}
\end{figure}

\FloatBarrier

\section{Background correction of intensity autocorrelation measurements}

We conduct intensity autocorrelation measurements using Hanbury-Brown Twiss interferometry. In these experiments, the fluorescence collection fibre is connected to a 50:50 fibre beamsplitter, with each end coupled into a single-photon avalanche photodiode. We include the effect of the background photoluminescence by renormalizing the $g^{(2)}(t)$ trace based on the parameter $p$ such that 

\begin{equation}
    g_p^{(2)}(t) = \frac{g^{(2)}(t) - (1-p^2)}{p^2}
\end{equation}

\noindent where $p$ is the fraction of total PL coming from the emitter \cite{Fishman2023}. We estimate $p$ from confocal scans of the emitter.

\newpage
\section{Kinetic model of the system}

We build a model with the goal of simulating the photoluminescence (PL) over time in various time-resolved and pulsed-microwave experiments. In order to do this, we build a set of rate equations describing the time-evolution of population of each level, and use this to simulate the PL  

\begin{equation}
    \label{SIEq:1}
    \dot{\boldsymbol\rho} = \textbf{M}(P_{\text{opt}},P_{\text{MW}}) \boldsymbol\rho
\end{equation}

\noindent where $\textbf{M}$ is the matrix describing the transfer of population from level $j$ into level $i$, and this is a function of the optical power $P_{\text{opt}}$ and the microwave power $P_{\text{MW}}$. In the basis given by $\chi = \{\boldsymbol\rho_{G}, \boldsymbol\rho_{E}, \boldsymbol\rho_{S0} \}$, where $\boldsymbol\rho_{G} = \{\rho_{Gz}, \rho_{Gy}, \rho_{Gx}\}$ and $\boldsymbol\rho_{E} = \{\rho_{Ez}, \rho_{Ey}, \rho_{Ex}\}$, this can be written as 

\begin{equation}
    \begin{bmatrix}
            \dot{\boldsymbol\rho_{G}} \\
            \dot{\boldsymbol\rho_{E}} \\
            \dot{\boldsymbol\rho_{S0}} \\
        \end{bmatrix} = 
        \begin{bmatrix}
            -\boldsymbol\Gamma_{G} + \boldsymbol\Gamma_{T_1} + \boldsymbol\Omega_{\text{MW}}(P_{\text{MW}}) & \boldsymbol\Gamma_{E\rightarrow G} & \textbf{k}_{S0 \rightarrow G}\\
            \boldsymbol\Gamma_{G\rightarrow E}(P_{\text{opt}}) & -\boldsymbol\Gamma_E & \textbf{0}\\
           \textbf{0} & \textbf{k}_{E \rightarrow S0} & -\boldsymbol\Gamma_{S0}\\
        \end{bmatrix}
        \begin{bmatrix}
            \boldsymbol\rho_{G} \\
            \boldsymbol\rho_{E} \\
            \boldsymbol\rho_{S0} \\
        \end{bmatrix}
\end{equation}

\noindent with 

\begin{equation}
\begin{split}
    &\boldsymbol\Gamma_{T_1} = \gamma_{T_1} \begin{bmatrix}
            0 & 1 & 1 \\
            1 & 0 & 1 \\
            1 & 1 & 0 \\
        \end{bmatrix} \\
    &\boldsymbol\Omega_{\text{MW}}(P_{\text{MW}}) = \Omega_{\text{MW}}(P_{\text{MW}}) \begin{bmatrix}
            0 & 1 & 0 \\
            1 & 0 & 0 \\
            0 & 0 & 0 \\
        \end{bmatrix} \\
    &\boldsymbol\Gamma_{E \rightarrow G} = \Gamma_{E \rightarrow G} \begin{bmatrix}
            1 & 0 & 0 \\
            0 & 1 & 0 \\
            0 & 0 & 1 \\
        \end{bmatrix} \\
    \end{split}
\end{equation}

\begin{equation}
    \begin{split}
    &\boldsymbol\Gamma_{G \rightarrow E}(P_{\text{opt}}) = \Gamma_{G \rightarrow E}(P_{\text{opt}}) \begin{bmatrix}
            1 & 0 & 0 \\
            0 & 1 & 0 \\
            0 & 0 & 1 \\
        \end{bmatrix} \\
    &\textbf{k}_{S0 \rightarrow G} = \begin{bmatrix}
            k_{S0 \rightarrow Gz} \\
            k_{S0 \rightarrow Gy} \\
            k_{S0 \rightarrow Gx} \\
        \end{bmatrix} \\
    &\textbf{k}_{E \rightarrow S0} = \begin{bmatrix}
            k_{Ez \rightarrow S0} & k_{Ey \rightarrow S0} & k_{Ex \rightarrow S0} \\
        \end{bmatrix}\\
    &\boldsymbol\Gamma_{G} = \begin{bmatrix}
            2 \gamma_{T_1} + \Gamma_{G \rightarrow E} (P_{\text{opt}}) + \Omega_{\text{MW}} (P_{\text{MW}}) & 0 & 0 \\
            0 & 2 \gamma_{T_1} + \Gamma_{G \rightarrow E} (P_{\text{opt}}) + \Omega_{\text{MW}} (P_{\text{MW}}) & 0 \\
            0 & 0 & 2 \gamma_{T_1} + \Gamma_{G \rightarrow E} (P_{\text{opt}}) \\
        \end{bmatrix} \\
    &\boldsymbol\Gamma_{E} = \begin{bmatrix}
            \Gamma_{E\rightarrow G} + k_{E0 \rightarrow S0} & 0 & 0 \\
            0 &  \Gamma_{E \rightarrow G} + k_{E+ \rightarrow S0} & 0 \\
            0 & 0 &  \Gamma_{E \rightarrow G} + k_{E- \rightarrow S0}  \\
        \end{bmatrix} \\
    &\boldsymbol\Gamma_{S0} = \begin{bmatrix} 
            k_{S0 \rightarrow Gz} + k_{S0 \rightarrow Gy} + k_{S0 \rightarrow Gx}\\
        \end{bmatrix} \\
\end{split}
\end{equation}

\noindent with rates defined in Fig.~2a of the main text. 

We get the population as a function of time by solving this set of coupled differential equations. In order to do this in a computationally inexpensive way, we can rewrite 

\begin{equation}
    \textbf{M} = \textbf{U}_{P_\text{opt},P_\text{MW}} \boldsymbol\lambda_{P_\text{opt},P_\text{MW}} \textbf{U}_{P_\text{opt},P_\text{MW}}^{-1}
\end{equation}

\noindent where $\textbf{U}_{P_\text{opt},P_\text{MW}}$ is the set of normalized eigenvectors of $\textbf{M}(P_\text{opt},P_\text{MW})$, and $\boldsymbol\lambda_{P_\text{opt},P_\text{MW}}$ is the diagonal matrix with eigenvalues of $\textbf{M}(P_\text{opt},P_\text{MW})$. Using this, we assume a solution to Eq.~\ref{SIEq:1} of the form

\begin{equation}
    \label{SIEq:solution}
    \boldsymbol\rho(t) = \textbf{U}_{P_\text{opt},P_\text{MW}} e^{(\boldsymbol\lambda_{P_\text{opt},P_\text{MW}} t)} \textbf{U}_{P_\text{opt},P_\text{MW}}^{-1} \boldsymbol\rho(t_0)
\end{equation}

In this way, the distribution of population over time is fully defined by the initial state $\boldsymbol\rho(t_0)$ and the set of rates included in $\textbf{M}(P_\text{opt},P_\text{MW})$. The eigenstate of $\textbf{M}(P_\text{opt},P_\text{MW})$ with eigenvalue $\lambda = 0$ gives the steady state population under a certain driving condition, $\boldsymbol\rho_{\text{ss}}(P_{\text{opt}},P_{\text{MW}})$. Finally, for a given population distribution, we assume that the photoluminescence is proportional to the radiative relaxation rate of each excited state sublevel times its population, such that

\begin{equation}
    \label{SIEq:PL}
    \text{PL}(t) = \sum \boldsymbol\Gamma_{E\rightarrow G} \boldsymbol\rho_{E} (t)
\end{equation}

We use this algorithm to calculate the population evolution -- and resulting photoluminescence -- of various experiments.

\subsection{$g^{(2)}(t)$ experiments}

To simulate the time-dependence observed in $g^{(2)}(t)$ experiments, we solve for 

\begin{equation}
    \label{SIEq:g2solution}
    \boldsymbol\rho_{g^{(2)}}(t) = \textbf{U}_{P_\text{opt},0} e^{(\boldsymbol\lambda_{P_\text{opt},0} t)} \textbf{U}_{P_\text{opt},0}^{-1} \boldsymbol\rho_{g^{(2)}}(0)
\end{equation}

\noindent where $\boldsymbol\rho_{g^{(2)}}(0)$ is the initial state of the system immediately after it emits a photon. This initial state is given by \cite{Exarhos2019}

\begin{equation}
\label{SIEq:x0g2}
    \boldsymbol\rho_{g^{(2)}}(0) =  \begin{bmatrix}
            \textbf{0} & \frac{\boldsymbol\Gamma_{E\rightarrow G}}{\textbf{e}^\top \boldsymbol\Gamma_{E\rightarrow G} \textbf{e}}& \textbf{0}\\ 
            \textbf{0} & \textbf{0} & \textbf{0}\\ 
            \textbf{0} & \textbf{0} & \textbf{0}\\ 
            \end{bmatrix}
            \begin{bmatrix}
            0 \\ \frac{\boldsymbol\rho_{\text{ss},E}(P_{\text{opt}},0)}{\textbf{e}^\top \boldsymbol\rho_{\text{ss},E}(P_{\text{opt}},0)} \\ 0 \\
        \end{bmatrix}
\end{equation}

\noindent where $\textbf{e}$ is a $3\times1$ vector with 1 at all entries. The probability of collecting a photon at a time $t$ after the initial photon detection is determined by the system PL at that time. In a $g^{(2)}(t)$ experiment, we are measuring the probability of detecting a photon a time $t$ after an initial photon emission, normalized by the unconditional probability of detecting a photon. Thus, the simulated $g^{(2)}(t)$ curve will be given by 

\begin{equation}
    \label{SIEq:g2Calc}
    g^{(2)}(t) = \frac{\sum \boldsymbol\Gamma_{E\rightarrow G}\boldsymbol\rho_{g^{(2)},E} (t)}{\sum \boldsymbol\Gamma_{E\rightarrow G}\boldsymbol\rho_{ss,E}(P_{\text{opt}},0) }
\end{equation}

\noindent where we have added a normalization factor corresponding to the probability of measuring a photon if the system is in the steady state. 

\subsection{Spin-dependent initialisation experiment}

In order to simulate the spin-dependent initialisation of the system (Fig.~2c of main text) we investigate the relative change in PL intensity a time $\tau_\text{r}$ after the start of an optical pulse due to the presence of a microwave pulse. We thus study the time-evolution of the population in the presence of optical drive in a reference experiment (without MW drive) and in a signal experiment (with a MW pulse between two subsequent optical pulses). 

\begin{figure*}[h]
  \includegraphics[width=0.5\textwidth]{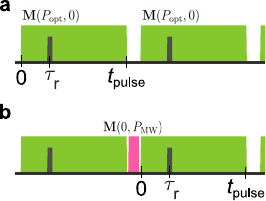}
  \caption{\textbf{Spin-dependent initialisation}. 
  \textbf{a} Reference and \textbf{b} signal sequences. Green blocks represent optical pulses, pink block represents a microwave $\pi$ pulse, and gray blocks represent a readout interval during which photons are collected. The readout interval is scanned across the duration of the optical pulse. 
  \label{SIFig::PLDecay}}
\end{figure*}

Figure~\ref{SIFig::PLDecay} shows the reference (a) and signal (b) pulse sequences, with relevant times specified. The optical pulse (green block) is chosen to be long (100s of microseconds) such that at $t_\text{pulse}$ the system has settled into the steady state $\boldsymbol\rho_{\text{ss}}(P_{\text{opt}},0)$ given by the eigenstate of $\textbf{M}(P_{\text{opt}},0)$ $\textbf{M}(0,P_{\text{MW}})$with eigenvalue $\lambda = 0$. We assume that the optical drive is weak, such that at any time the population in the excited state is only a small fraction of the entire population. Since the time between the two optical pulses is small ($\sim 100$~ns), we can assume that, for the reference experiment, the initial state $\boldsymbol\rho_{\text{ref}}(0)$ is given by $\boldsymbol\rho_{\text{ss}}(P_{\text{opt}},0)$. In contrast, for the signal experiment we assume that the initial state is given by $\boldsymbol\rho_{\text{sig}}(0) = \boldsymbol\Pi \boldsymbol\rho_{\text{ss}}(P_{\text{opt}},0)$, where $\boldsymbol\Pi$ is the operator that swaps $\rho_{Gz}$ and $\rho_{Gy}$. Using these expressions for $\boldsymbol\rho_{\text{ref}}(0)$ and $\boldsymbol\rho_{\text{sig}}(0)$, we then calculate the population dependence on $\tau_\text{r}$, 

\begin{equation}
    \begin{split}
    \boldsymbol\rho_\text{ref}(\tau_\text{r}) = \textbf{U}_{P_\text{opt},0} e^{(\boldsymbol\lambda_{P_\text{opt},0} \tau_\text{r})} \textbf{U}_{P_\text{opt},0}^{-1} \boldsymbol\rho_{\text{ref}}(0) \\
    \boldsymbol\rho_\text{sig}(\tau_\text{r}) = \textbf{U}_{P_\text{opt},0} e^{(\boldsymbol\lambda_{P_\text{opt},0} \tau_\text{r})} \textbf{U}_{P_\text{opt},0}^{-1} \boldsymbol\rho_{\text{sig}}(0) \\
    \end{split}
\end{equation}

\noindent and combine this with Eq.~\ref{SIEq:PL} to obtain the contrast as a function of $\tau_\text{r}$:

\begin{equation}
    \label{SIEq:PLCalc}
    C_{\text{PL}}(\tau_\text{r}) = \frac{\sum (\boldsymbol\Gamma_{E\rightarrow G} \boldsymbol\rho_{\text{sig},E}(\tau_\text{r}) -\boldsymbol\Gamma_{E\rightarrow G} \boldsymbol\rho_{\text{ref},E}(\tau_\text{r}))}{ \sum (\boldsymbol\Gamma_{E\rightarrow G} \boldsymbol\rho_{\text{ref},E}(\tau_\text{r}))}
\end{equation}

\subsection{Modified spin-relaxation experiment}

\begin{figure*}[h]
  \includegraphics[width=0.5\textwidth]{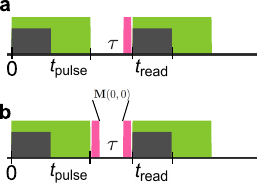}
  \caption{\textbf{Modified spin relaxation experiment}. 
  \textbf{a} Reference and \textbf{b} signal sequences. Green blocks represent optical pulses, pink blocks represents a microwave $\pi$ pulse, and gray blocks represent a readout interval during which photons are collected. The time interval between the two microwave pulses in the signal sequence, indicated by $\tau$, is scanned during the experiment. 
  \label{SIFig::T1}}
\end{figure*}

In order to simulate the behavior of the system in a modified spin-relaxation experiment (Fig.~2d of main text) we proceed similarly as above, but now investigate the dependence of a variable interval of length $\tau$ during the pulses. Figure~\ref{SIFig::PLDecay} shows the reference (a) and signal (b) pulse sequences, with relevant times specified.

At $t=t_\text{pulse}$, i.e. at the end of a long optical pulse, the system is in state $\boldsymbol\rho_{\text{ss}}(P_{\text{opt}},0)$. In the reference experiment, this is followed by a variable delay $\tau$, where the population evolves as determined by the rate matrix $\textbf{M}(0,0)$. In the signal experiment, the time delay $\tau$ is preceded by a microwave pulse represented by the operator $\boldsymbol \Pi$. In both reference and signal experiments, a microwave pulse is applied after the delay $\tau$ and just before the arrival of the optical pulse. The state of the system at $t=0$, when the optical pulse arrives, is given by 

\begin{equation}
    \begin{split}
    &\boldsymbol\rho_{\text{ref}}(0) = \boldsymbol \Pi  (\textbf{U}_{0,0} e^{(\boldsymbol\lambda_{0,0} \tau)} \textbf{U}_{0,0}^{-1}) \boldsymbol\rho_{\text{ss}}(P_{\text{opt}},0)\\
    &\boldsymbol\rho_{\text{sig}}(0) = \boldsymbol\Pi  (\textbf{U}_{0,0} e^{(\boldsymbol\lambda_{0,0} \tau)} \textbf{U}_{0,0}^{-1}) \boldsymbol\Pi \boldsymbol\rho_{\text{ss}}(P_{\text{opt}},0)
    \end{split}
\end{equation}

The subsequent population evolution during the optical drive is then given by 

\begin{equation}
    \begin{split}
    \boldsymbol\rho_\text{ref}(t) = \textbf{U}_{P_\text{opt},0} e^{(\boldsymbol\lambda_{P_\text{opt},0} t)} \textbf{U}_{P_\text{opt},0}^{-1} \boldsymbol\rho_{\text{ref}}(0) \\
    \boldsymbol\rho_\text{sig}(t) = \textbf{U}_{P_\text{opt},0} e^{(\boldsymbol\lambda_{P_\text{opt},0} t)} \textbf{U}_{P_\text{opt},0}^{-1} \boldsymbol\rho_{\text{sig}}(0) \\
    \end{split}
\end{equation}

\noindent which we combine with Eq.~\ref{SIEq:PL} to calculate the PL during the optical drive pulse. We integrate this between $t=0$ and $t=t_\text{read}$ to obtain the integrated PL during the readout time lasting approximately $100~\mu$s:

\begin{equation}
    \label{SIEq:T1Calc}
    \begin{split}
        &C_{T_1}(\tau) = \frac{\text{PL}_\text{sig}-\text{PL}_\text{ref}}{\text{PL}_\text{ref}},\\
        &\text{PL}_\text{ref} = \int_0^{t_\text{read}} \sum \boldsymbol\Gamma_{E\rightarrow G} \boldsymbol\rho_\text{ref}(t) \, dt \\
        &\text{PL}_\text{sig} = \int_0^{t_\text{read}} \sum \boldsymbol\Gamma_{E\rightarrow G} \boldsymbol\rho_\text{sig}(t) \, dt 
    \end{split}
\end{equation}


\subsection{cwODMR contrast}

In order to calculate the cwODMR contrast, we compare the steady-state PL in the presence of simultaneous optical and microwave drives ($\boldsymbol\rho_{\text{ss}}(P_{\text{opt}},P_\text{MW})$) to the steady-state PL in the presence of optical drive and absence of microwave drive ($\boldsymbol\rho_{\text{ss}}(P_{\text{opt}},0)$). These are respectively determined by the eigenstates of $\textbf{M}(P_{\text{opt}},P_{\text{MW}})$ and $\textbf{M}(P_{\text{opt}},P_{\text{MW}}=0)$ with eigenvalues equal to zero. This gives, for the cwODMR contrast, 

\begin{equation}
        C_{\text{cw}} = \frac{\sum (\boldsymbol\Gamma_{E\rightarrow G} \boldsymbol\rho_{\text{ss},E}(P_{\text{opt}},P_\text{MW}) - \boldsymbol\Gamma_{E\rightarrow G} \boldsymbol\rho_{\text{ss},E}(P_{\text{opt}},0))}{ \sum \boldsymbol\Gamma_{E\rightarrow G} \boldsymbol\rho_{\text{ss},E}(P_{\text{opt}},0)}
\end{equation}


\subsection{Fitting Procedure}

In order to determine the parameters presented in the schematics of Fig.~2a of the main text for our system, we fit the predictions of the model presented in the section above to the experimental results presented in Fig.~2b-d of the main text. We do this by minimizing the error given by 

\begin{equation}
        \delta_\text{total} = \delta_{g^{(2)}} + \delta_{\text{PL}} + \delta_{T_1}
\end{equation}

\noindent where 

\begin{equation}
\begin{split}
    &\delta_{g^{(2)}} = \sum_{t_i} \tfrac{1}{N} (g^{(2)}_\text{calc} (t_i) - g^{(2)}_\text{exp} (t_i))^2,\\
    &\delta_{\text{PL}} = \sum_{\tau_{\text{r},i}} \tfrac{1}{N} (A_\text{PL} C_\text{PL,calc}(\tau_{\text{r},i}) - C_\text{PL,exp}(\tau_{\text{r},i}))^2, \\
    &\delta_{T_1} = \sum_{\tau_i} \tfrac{1}{N} (A_{T_1} C_{T_1,\text{calc}}(\tau_i)  - C_{T_1,\text{exp}}(\tau_i))^2,
\end{split}
\end{equation}

\noindent with $g^{(2)}_\text{calc} (t)$, $C_\text{PL,calc}(t)$ and $C_{T_1,\text{calc}}(t)$ defined in Eqs.~\ref{SIEq:g2Calc}, \ref{SIEq:PLCalc} and \ref{SIEq:T1Calc}, respectively. The subindices calc and exp refer respectively to calculated or experimentally observed quantities, and the $\tfrac{1}{N}$ factor normalizes for the number of observations made in each measurement. The prefactors $A_\text{PL}$ and $A_{T_1}$ are phenomenological scaling factors between 0 and 1 that account for the fact that we often observe lower contrast in pulsed than cw microwave experiments, a feature that could arise from imperfect microwave delivery due to impedance mismatches in our microwave line. Finally, we constrain the fit to parameter combinations that provide calculated cwODMR contrast at saturated microwave drive condition equal to or higher to the measured cwODMR contrast. We obtain the best estimates for the parameters $\{ \Gamma_{E\rightarrow G}, \Gamma_{G \rightarrow E}, k_{Ex,z \rightarrow S0}, k_{Ey \rightarrow S0}, k_{S0 \rightarrow Gx,z}, k_{S0 \rightarrow Gy}, \gamma_{T_1}, A_\text{PL}, A_{T_1} \}$ from this minimization procedure. We assume here that $k_{Ex,z \rightarrow S0} = k_{Ex \rightarrow S0} = k_{Ez \rightarrow S0}$, and analogously for the reverse intersystem crossing rates in order to minimize the number of fit parameters involved.

\FloatBarrier
\newpage
\section{Comparison of ground-state and metastable spin triplet models}

\begin{figure}[h!]
    \centering
    \includegraphics[width=1\linewidth]{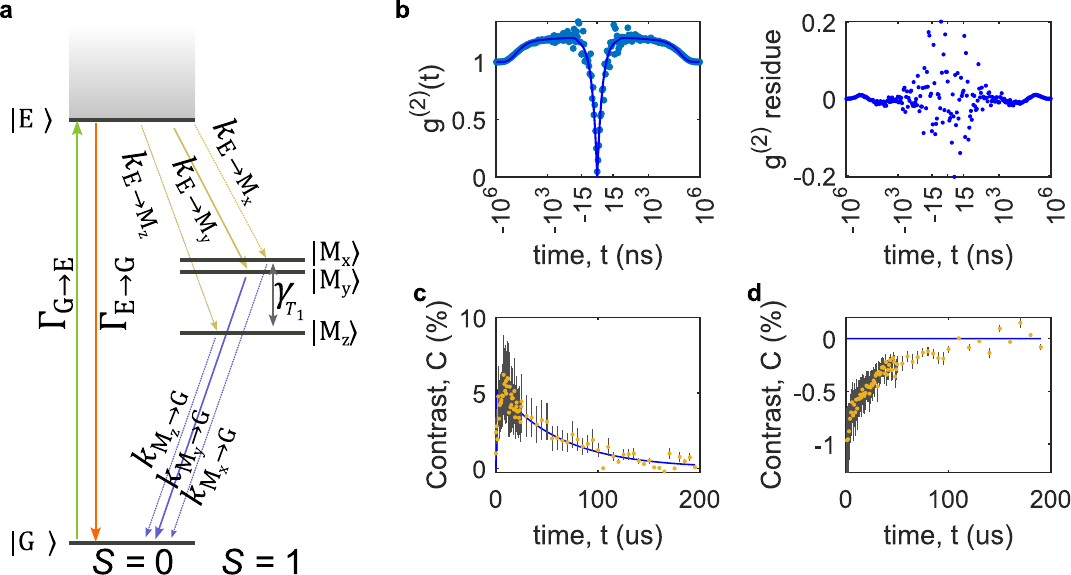}
    \caption{\textbf{Model with a spin-triplet in the metastable state, and spin singlet ground and optically excited states.}  
    \textbf{a} Description of the model and relevant rates; \textbf{b} background-corrected results of an intensity autocorrelation experiment (blue circles), with fit result (blue curve) and residuals of the fit (right panel). \textbf{c} Experimental results (yellow circles) of the spin-dependent initialisation experiment and \textbf{d} modified $T_1$ experiment, accompanied by results of a fit of the model to the experimental data (blue curves). \label{SIFig::singlet}}
\end{figure}

We consider a model consisting of spin singlet ground and excited states, and a triplet metastable state, reminiscent of what is seen in the case of organic molecules like pentacene and the ST1 and TR12 defects in diamond \cite{Mena2024, Balasubramanian2019,Foglszinger2022}. We consider a model as presented in Fig.~\ref{SIFig::singlet}, and proceed to obtain a global fit of this model to the experimental results. The results of this fit are presented in Fig.~\ref{SIFig::singlet}. While the model is able to capture the behaviour observed in the intensity autocorrelation and the spin-dependent initialisation experiments, it is not able to capture the behavior of the modified $T_1$ experiment. 

\begin{figure}[h!]
    \centering
    \includegraphics[width=0.6\linewidth]{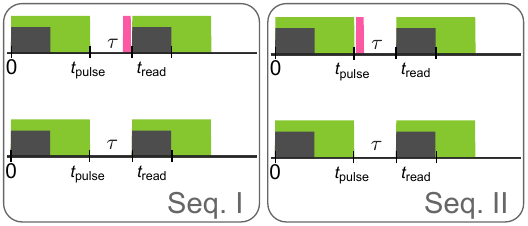}
    \caption{Pulse sequences used to measure the spin-dependent relaxation time. Green blocks represent optical pulses, pink blocks represents a microwave $\pi$ pulse, and gray blocks represent a readout interval during which photons are collected. Sequences on the top indicate signal sequence, whereas the sequences in the bottom indicate the reference sequences. The time interval before (Seq. I) and after (Seq. II) the microwave pulse in the signal sequence, indicated by $\tau$, is scanned during the experiment.  \label{SIFig::T1BeforeAfter}}
    
\end{figure}

In addition, we measure the spin-relaxation of different defects using two different pulse sequences as indicated in Fig.~\ref{SIFig::T1BeforeAfter}. These sequences correspond to two different experiments probing the spin-dependent relaxation, where the contrast-inducing microwave $\pi$ pulse on resonance with a microwave resonance of the system occurs after (Seq. I) or before (Seq. II) a variable time $\tau$. For a system with a spin-triplet in the metastable state, the spin-dependent pulsed-ODMR (pODMR) contrast decay in these experiments is expected to reflect the differences in spin-dependent reverse intersystem crossing rates \cite{Mena2024}. In this case, we expect to observe relaxation times that differ significantly depending on whether the $\pi$ pulse occurs before or after a wait time. In contrast, in the case of a ground-state spin triplet, the relaxation dynamics from either spin sublevel is expected to be similar and dominated by spin-lattice relaxation. 

We perform this experiment on two different emitters respectively at room temperature and at 4~K, with results presented in Fig.~\ref{fig:T1pipulseanalysis}. We observe that the two sequences result in equal relaxation timescales for the decay of pODMR contrast. These results further support our assignment for the configuration of energy levels for this defect type, with a ground-state spin triplet and a metastable state spin singlet. 

\begin{figure}[h!]
    \centering
    \includegraphics[width=\linewidth]{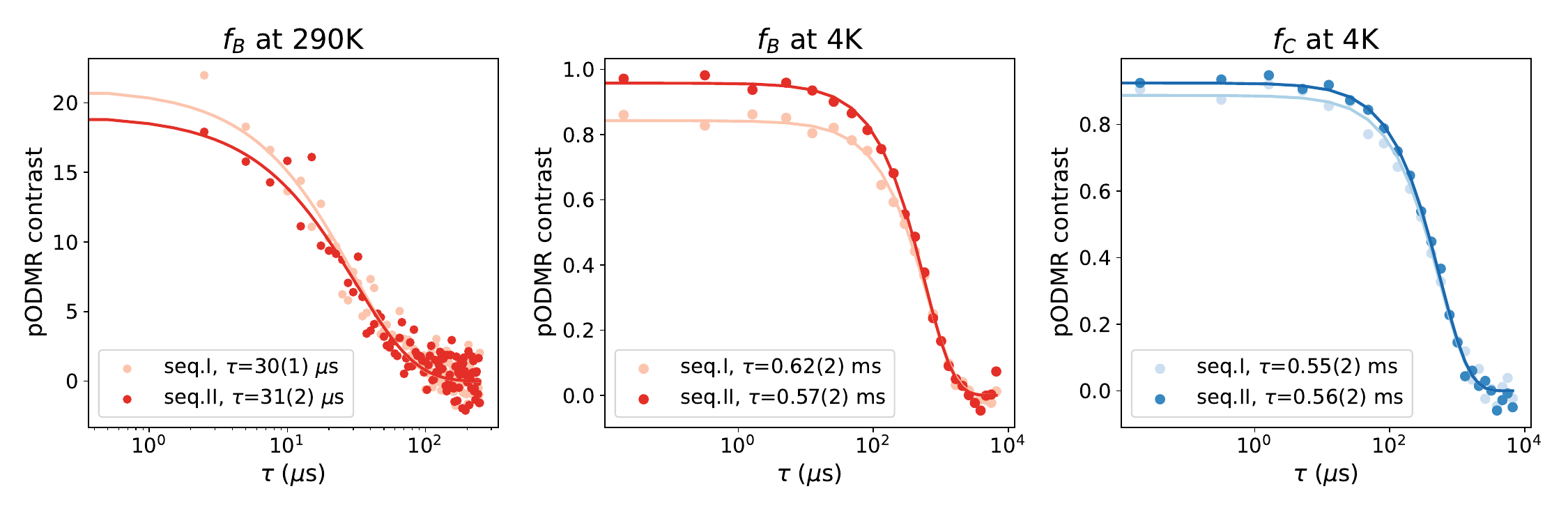}
    \caption{\textbf{Results of a spin-relaxation experiment} where the contrast-inducing microwave $\pi$ pulse occurs after (Seq. I) or before (Seq. II) a variable delay time $\tau$. Left panel presents the result of this experiment on a defect at room-temperature, and when the $\pi$ pulse is on resonance with \textit{f}\textsubscript{B}. The middle and right panels present the result of this experiment on a defect at 4~K, and with the $\pi$ pulse on resonance with \textit{f}\textsubscript{B} and \textit{f}\textsubscript{C}, respectively.\label{fig:T1pipulseanalysis}} 
\end{figure}

\FloatBarrier
\newpage
\section{pODMR and $g^2(t)$ data with associated global fits for additional defects}

\begin{table}
    \centering
    \caption{\textbf{Model parameters.} Summary of key parameters obtained from fitting the data in Figs.~\ref{SIFig::Em4}-\ref{SIFig::Em12} to the model with a triplet ground state (Main Text, Fig.~2a).}
    \label{Tab::tab1}
    \begin{tabular}{|c|c|c|c|c|c|c|c|c|c|c|}
        \hline
        Rate & $\Gamma_{G\rightarrow E}$ & $\Gamma_{E\rightarrow G}$ & $k_{E_x \rightarrow S0}$ & $k_{E_y \rightarrow S0}$ & $k_{E_z \rightarrow S0}$ & $k_{S0 \rightarrow G_x}$ & $k_{S0 \rightarrow G_y}$ &$k_{S0 \rightarrow G_z}$ & $\gamma_{T_1}$ & \textit{f}\textsubscript{B} Contrast\\
         Unit & kHz/$\mu$W & MHz & MHz & MHz & MHz & kHz & kHz & kHz  & kHz & \% \\
         \hline \hline
         Def4 & 0.18 &	118&	114&	220&	114&	26&	248 &	26&	1.0 & 3.7 \\
         Def3744 &0.07&	220 &	149&	456&	149&	633&	2661&	633&	8.3 & 4\\
         Def15 & 0.9	&143&	34&	388&	34&	5&	856&	5&	0.3 & 9.3\\
         Def3981 &0.92&	138&	10&	253&	10&	166&	5200&	166	&0.23  & 9.5\\
         DefJ19 &0.92 &	163.4&	5.4&	190&	5.4&	2&	675&	2&	3.2 & 22\\
         Def12 & 7.1 &168&	21&	529&	21&	35&	2605 &35&	1.3 & 30\\
         \hline
    \end{tabular}
\end{table}

\begin{figure}[h!]
    \centering
    \includegraphics[width=1\linewidth]{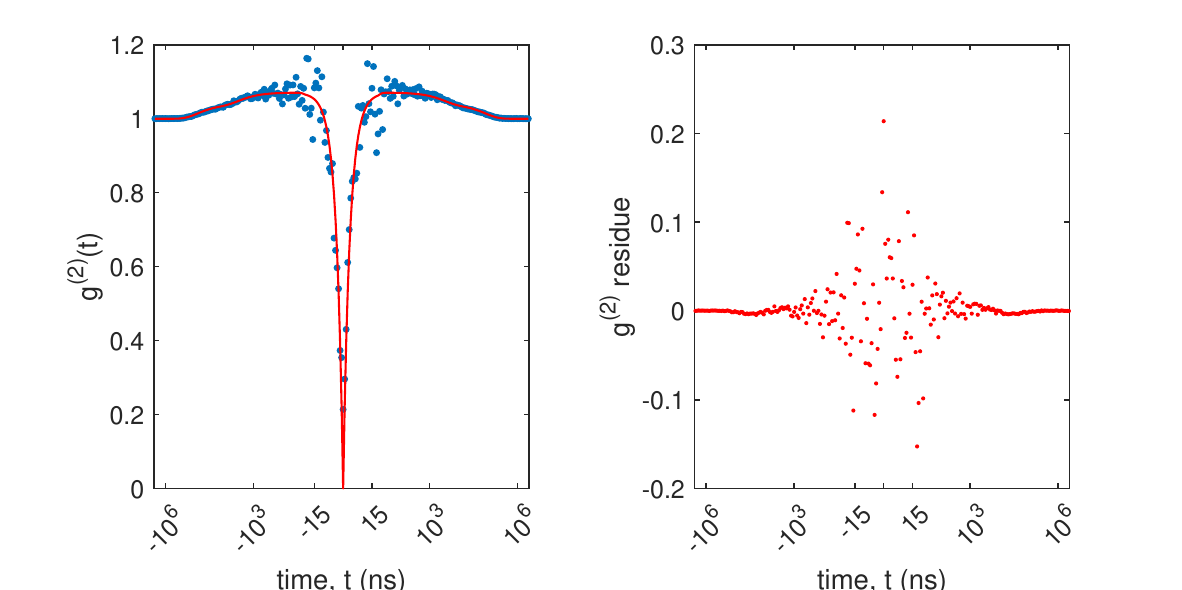}
    \caption{\textbf{Defect 4.} Background-corrected intensity autocorrelation used to extract rates in Tab.~\ref{Tab::tab1} (left) and residual of the fit (right). Background is typically not more than 10\% of emitter PL. Data is presented as blue circles, and result of the fit is presented as red curve. \label{SIFig::Em4}}
\end{figure}

\begin{figure}[h!]
    \centering
    \includegraphics[width=1\linewidth]{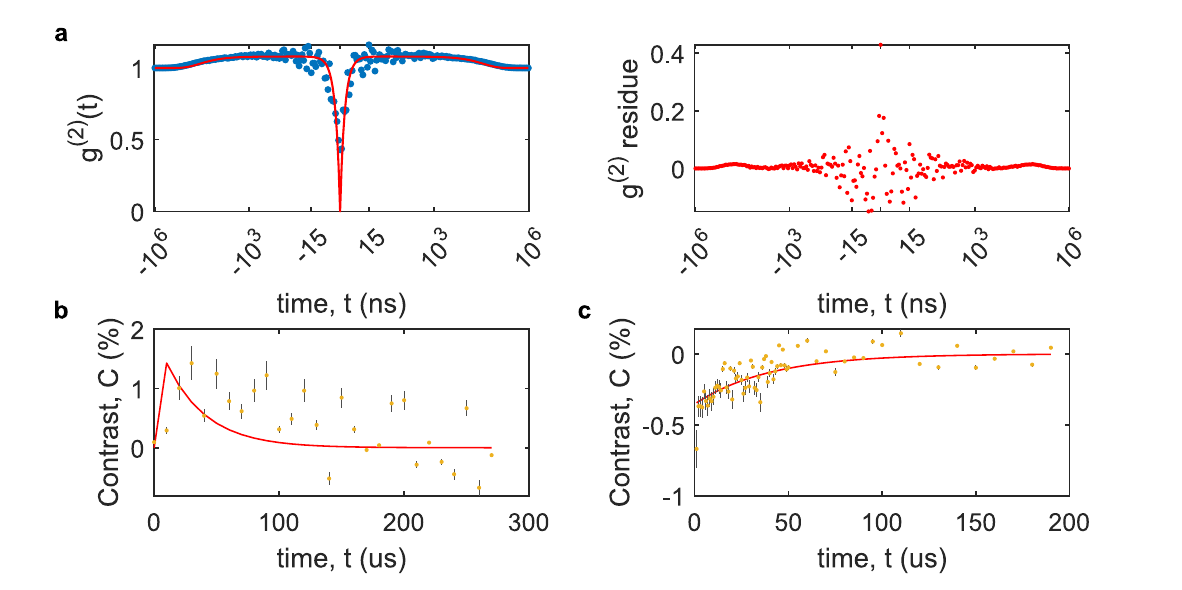}
    \caption{\textbf{Defect 3744.} Experimental data (blue and yellow circles) and results of fits (red curves) used to extract rates in Tab.~\ref{Tab::tab1}. \textbf{a} Background-corrected intensity autocorrelation (left) and residual of the fit of the intensity autocorrelation experiment (right). Background is typically not more than 10\% of emitter PL. \textbf{b} Spin-dependent initialisation, and \textbf{c} modified $T_1$. The error bars correspond to one standard deviation of the measured data. \label{SIFig::Em3744}}
\end{figure}

\begin{figure}[h!]
    \centering
    \includegraphics[width=1\linewidth]{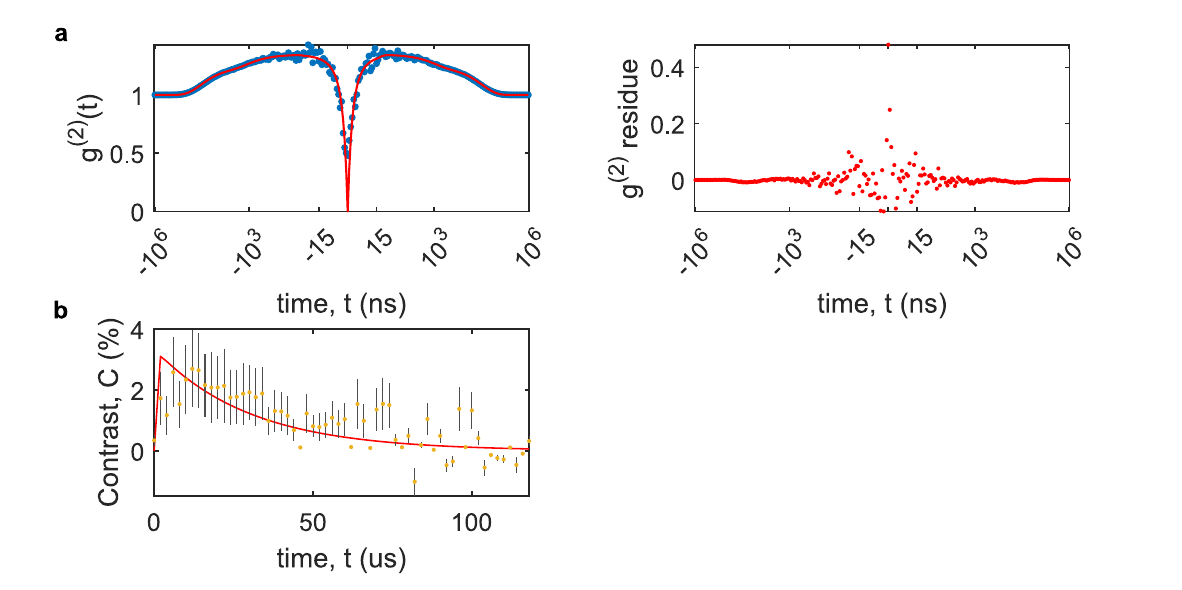}
    \caption{\textbf{Defect 15.}  Experimental data (blue and yellow circles) and results of fits (red curves) used to extract rates in Tab.~\ref{Tab::tab1}. \textbf{a} Background-corrected intensity autocorrelation (left) and residual of the fit of the intensity autocorrelation experiment (right). Background is typically not more than 10\% of emitter PL. \textbf{b} Spin-dependent initialisation. The error bars correspond to one standard deviation of the measured data. \label{SIFig::Em15}}
\end{figure}

\begin{figure}[h!]
    \centering
    \includegraphics[width=1\linewidth]{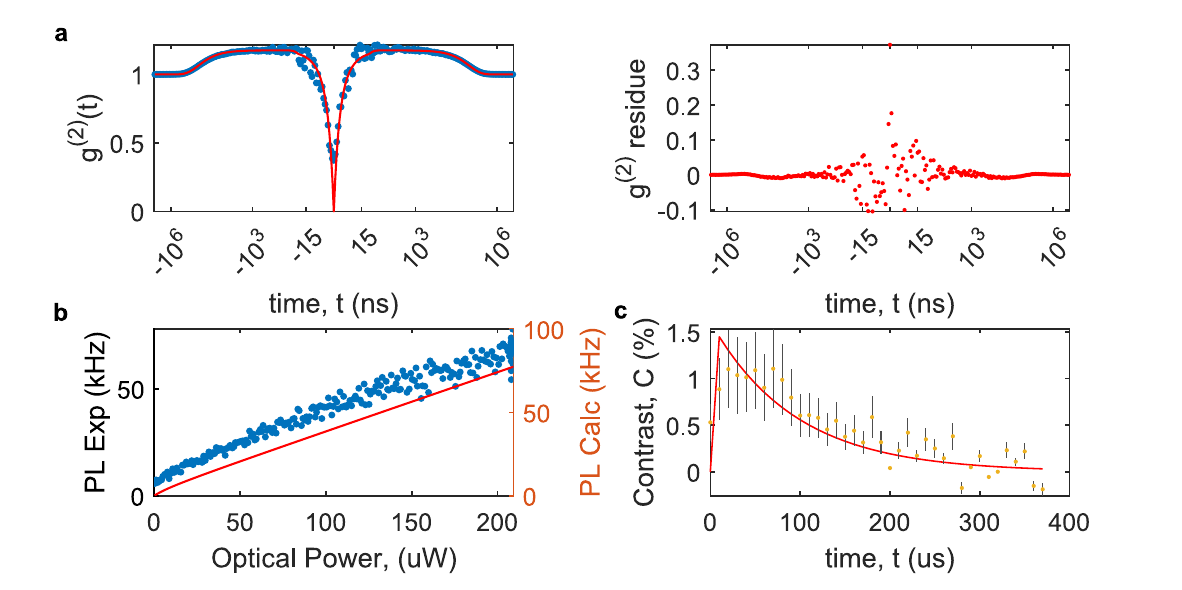}
    \caption{\textbf{Defect 3981.}  Experimental data (blue and yellow circles) and results of fits (red curves) used to extract rates in Tab.~\ref{Tab::tab1}. \textbf{a} Background-corrected intensity autocorrelation (left) and residual of the fit of the intensity autocorrelation experiment (right). Background is typically not more than 10\% of emitter PL. \textbf{b} PL saturation and \textbf{c} Spin-dependent initialisation. The error bars correspond to one standard deviation of the measured data. \label{SIFig::Em3981}}
\end{figure}

\begin{figure}[h!]
    \centering
    \includegraphics[width=1\linewidth]{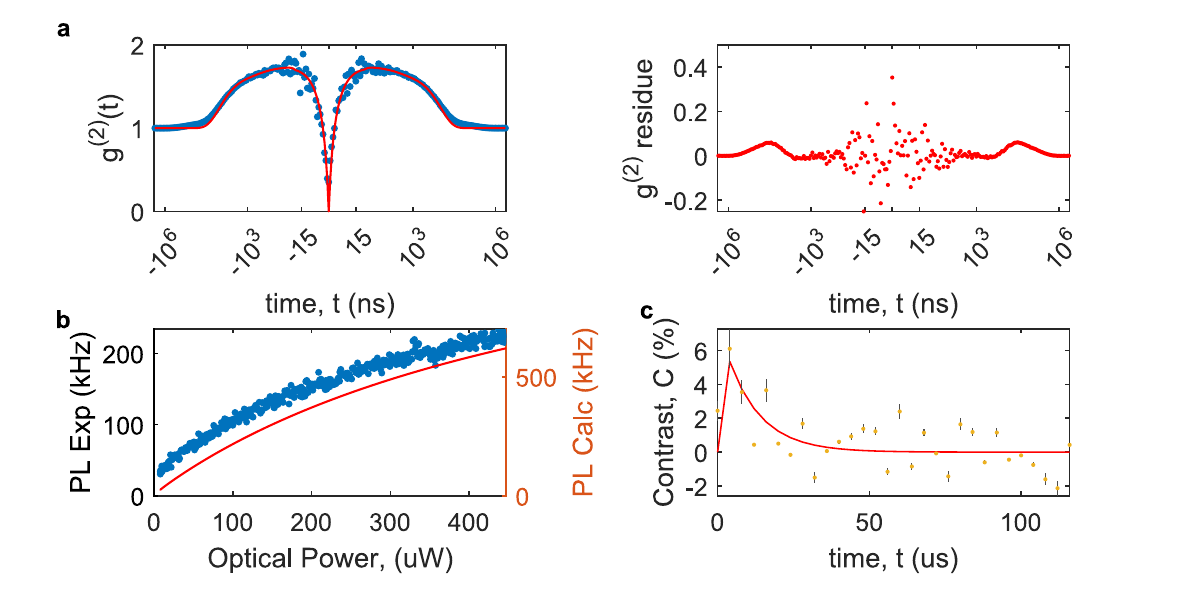}
    \caption{\textbf{Defect 12.}  Experimental data (blue and yellow circles) and results of fits (red curves) used to extract rates in Tab.~\ref{Tab::tab1}. \textbf{a} Background-corrected intensity autocorrelation (left) and residual of the fit of the intensity autocorrelation experiment (right). Background is typically not more than 10\% of emitter PL. \textbf{b} PL saturation and \textbf{c} Spin-dependent initialisation. The error bars correspond to one standard deviation of the measured data. \label{SIFig::Em12}}
\end{figure}

\FloatBarrier
\newpage

\section{cwODMR spectra}

\subsection{cwODMR spectra associated with Figure 2e}

\begin{figure}[h!]
    \centering
    \includegraphics[width=0.5\linewidth]{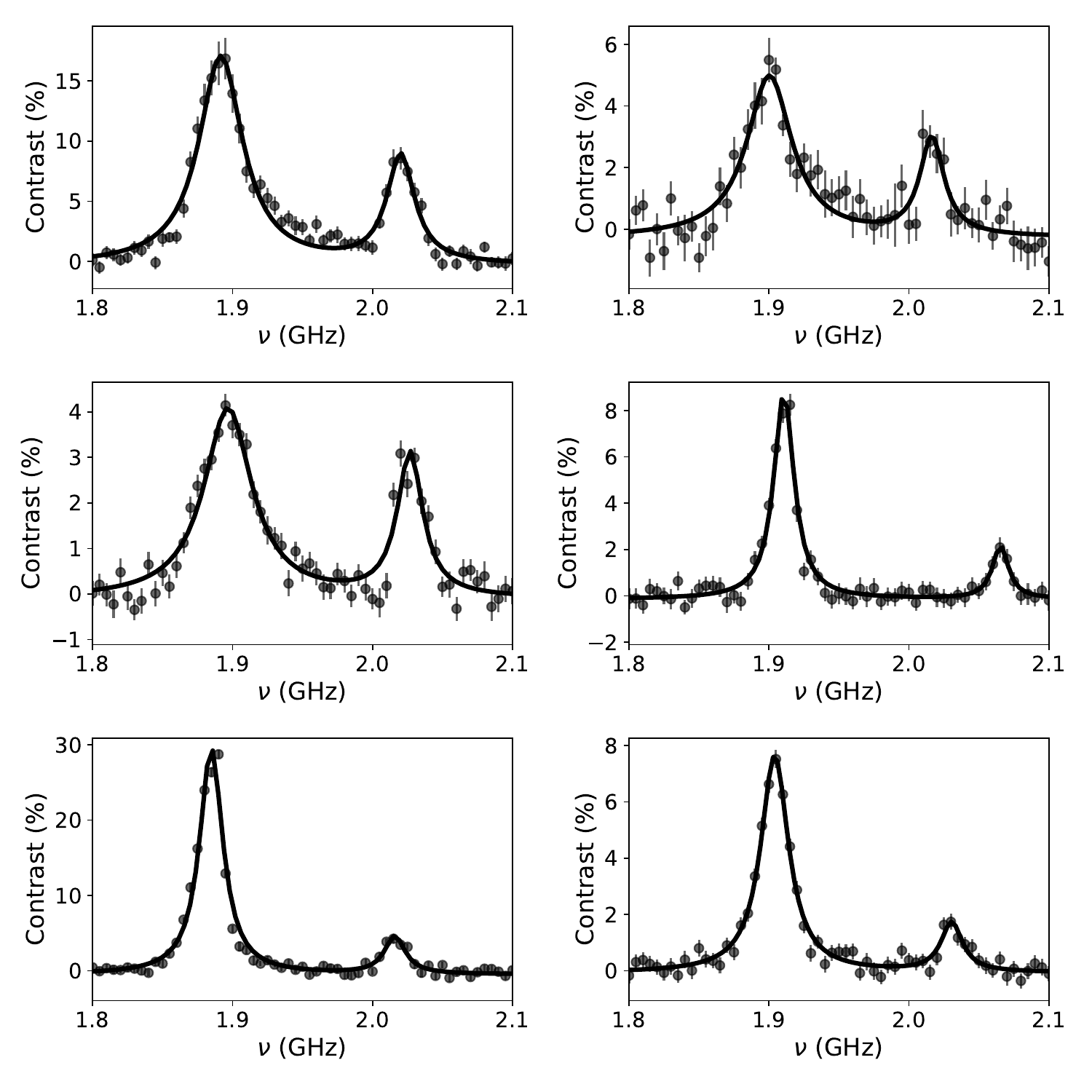}
    \caption{\textbf{cwODMR spectra associated with Figure 2e.} cwODMR Spectra taken under saturation conditions for different defects in the absence of magnetic field. The lower frequency peak corresponds to \textit{f}\textsubscript{B}, the higher frequency peak to \textit{f}\textsubscript{C}. The saturated contrast of each \textit{f}\textsubscript{B} transition is plotted in Fig.~2e. The error bars correspond to the standard error of the mean of the measured data.}
    \label{SIFig:2e_cwodmr}
\end{figure}

\subsection{cwODMR spectra associated with Figure 3b}

CW ODMR spectra under 50 mT field are presented below, with $\phi=90$, varying $\theta$. This corresponds to rotation in the $zy$ plane of the defect, such that $\theta=0, \phi=90$ is along $z$ and $\theta=90, \phi=90$ is along $y$. Physical setup constraints restrict measurements at angles outside of the range presented.
The left column (green) corresponds to the \textit{f}\textsubscript{A} transition, the middle column (red) to the \textit{f}\textsubscript{B} transition, and the right column (blue) to the \textit{f}\textsubscript{C} transition.

A Lorentzian fit to each spectrum is shown as a black curve, identifying the transition frequency for each resonance. In some cases, additional peaks are observed near \textit{f}\textsubscript{A}, which we assign to replicas of the expected peaks, as they match the half-frequency transitions of \textit{f}\textsubscript{A} and \textit{f}\textsubscript{B}. Their presence is due to the generation of second and third-order harmonics of the microwave carrier frequency at high output powers (1.5~mW into amplifier, 1.6~W after amplification). The replicas disappear at lower microwave powers ($<$0.1~W after amplification).

\begin{figure}[h!]
    \centering
    \includegraphics[width=0.7\linewidth]{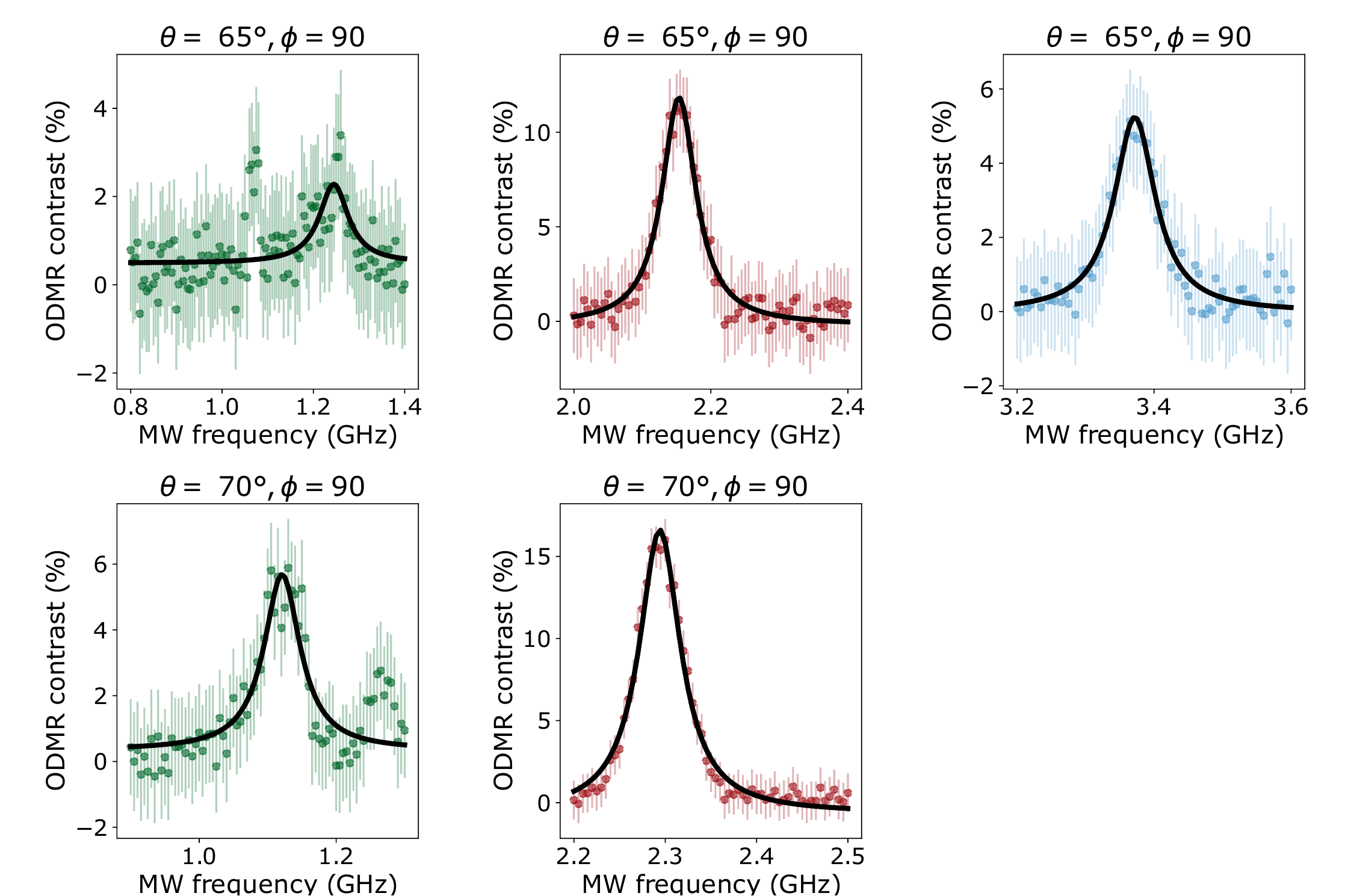}
    \caption{cwODMR spectra associated with Figure 3b at 50mT, with $\phi=90$, varying $\theta$. The error bars correspond to the standard error of the mean of the measured data.}
\end{figure}

\begin{figure}[h!]
    \centering
    \includegraphics[width=0.7\linewidth]{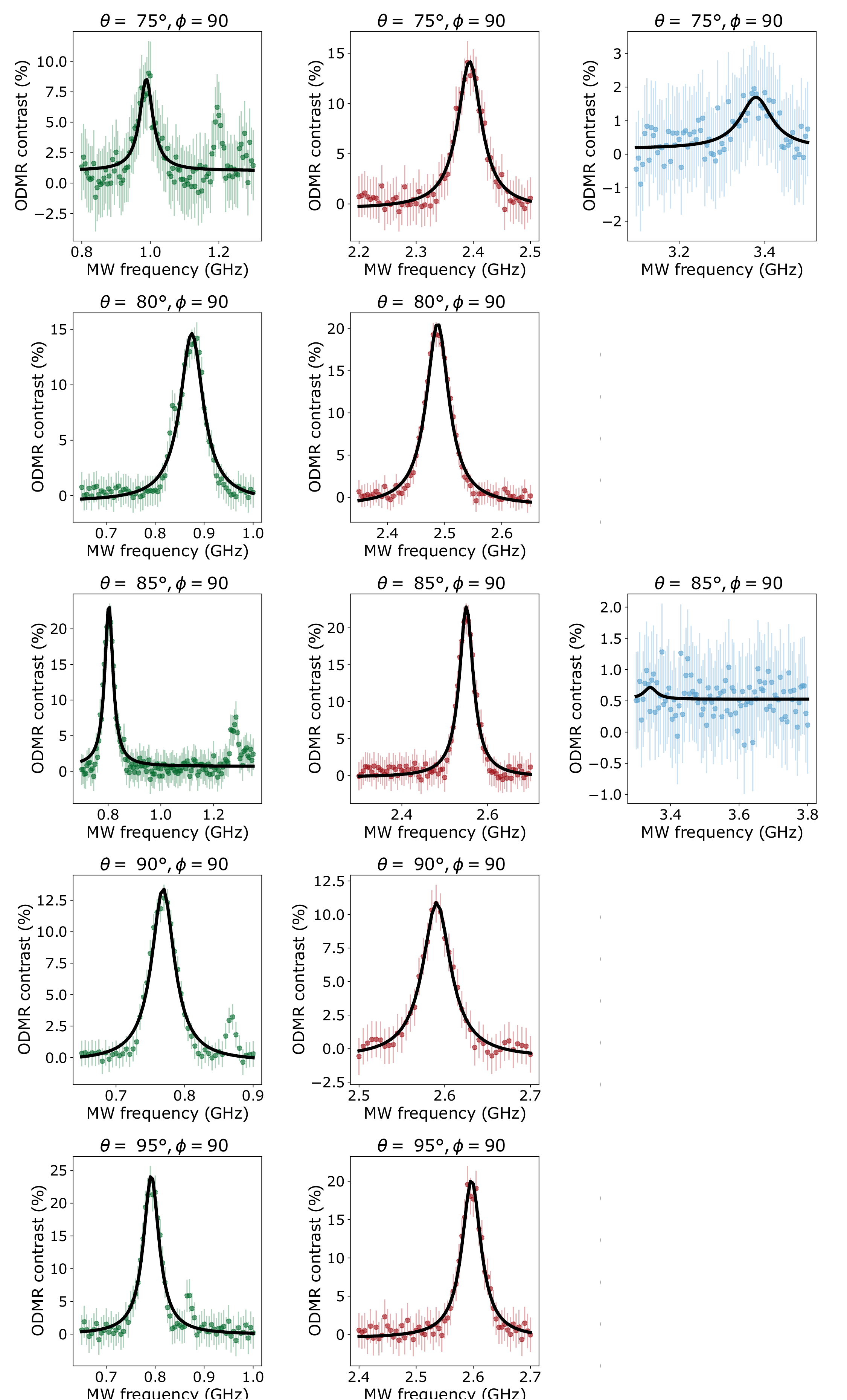}
    \caption{cwODMR spectra associated with Figure 3b at 50mT, with $\phi=90$, varying $\theta$. The error bars correspond to the standard error of the mean of the measured data.}
\end{figure}

\begin{figure}[h!]
    \centering
    \includegraphics[width=0.7\linewidth]{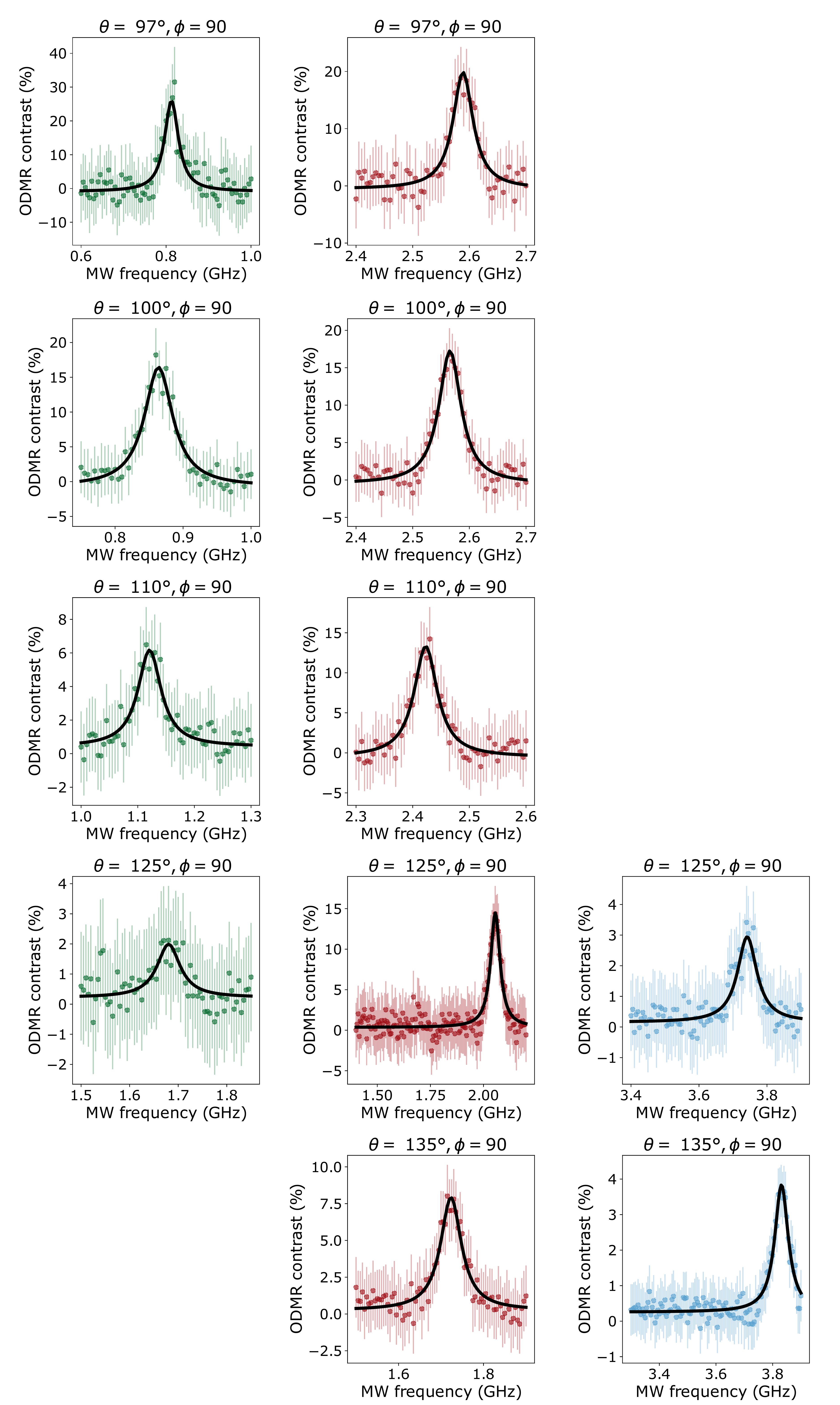}
    \caption{cwODMR spectra associated with Figure 3b at 50mT, with $\phi=90$, varying $\theta$.}
\end{figure}

\FloatBarrier
\newpage
\subsection{cwODMR spectra associated with Figure 3c}

CW ODMR spectra under 50 mT field are presented below, with varying $\phi$, $\theta=95$. This corresponds to rotation near the $xy$ plane of the defect, where the $x$ axis is given by $\phi=0, \theta=90$, and $y$ is along $\phi=90, \theta=90$. A Lorentzian fit to each spectrum is shown as a black curve, identifying the transition frequency for each resonance. As before, we observe replicas alongside expected peaks in the spectra.  Physical setup constraints restrict measurements at angles outside of the range presented. 
The left column (green) corresponds to the \textit{f}\textsubscript{A} transition and the right column (red) to the \textit{f}\textsubscript{B} transition.The contrast of \textit{f}\textsubscript{C} is not observable in this magnetic field orientation.

\begin{figure}[h!]
    \centering
    \includegraphics[width=0.6\linewidth]{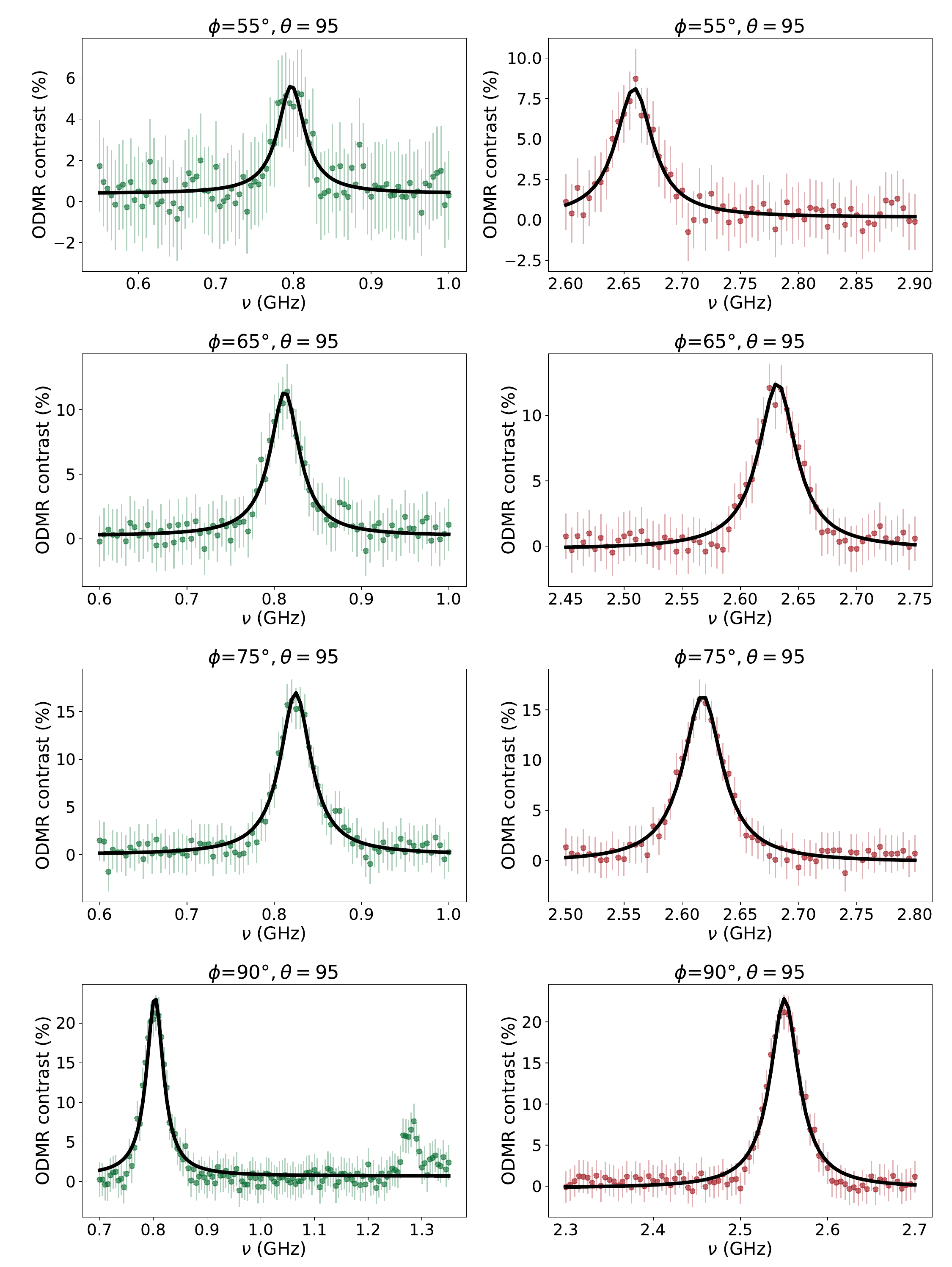}
    \caption{cwODMR spectra associated with Figure 3c at 50mT, with $\theta=95$, varying $\phi$.}
\end{figure}

\begin{figure}[h!]
    \centering
    \includegraphics[width=0.6\linewidth]{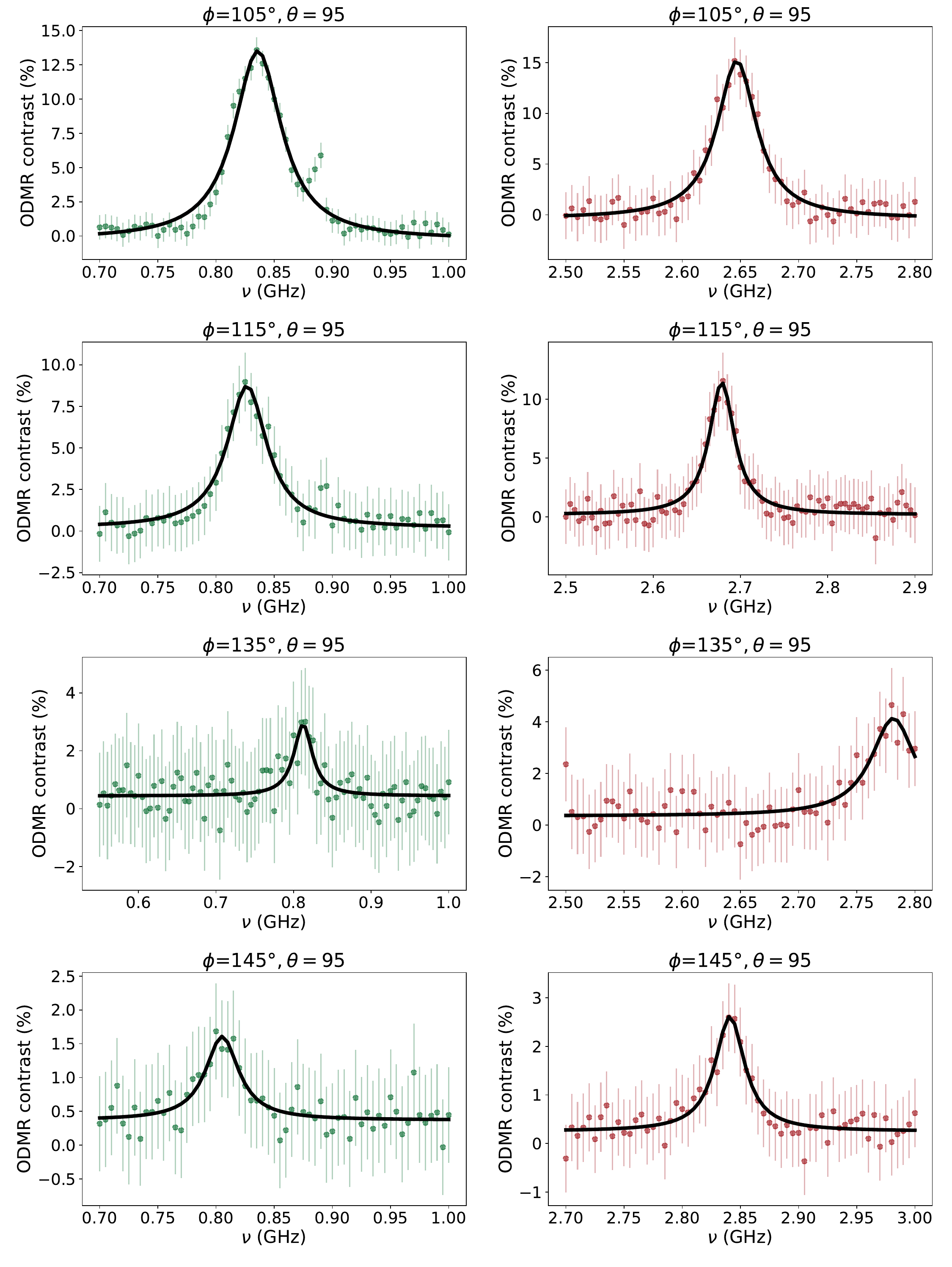}
    \caption{cwODMR spectra associated with Figure 3c at 50mT, with $\theta=95$, varying $\phi$.}
\end{figure}

\FloatBarrier

\newpage

\section{Excited state zero-field splitting parameters}

The excited-state zero-field splitting parameters, $D_\text{ES}$, $E_\text{ES}$ directly govern the effect of bias magnetic field in mixing the excited-state zero-field eigenstates, and therefore have direct influence on the spin-selectivity of the direct intersystem crossing rates. Thus, we would expect that the magnitude of $D_\text{ES}$, $E_\text{ES}$ would influence the spin-initialisation cycle at applied magnetic field.

\begin{figure}[h!]
    \centering
    \includegraphics[width=1\linewidth]{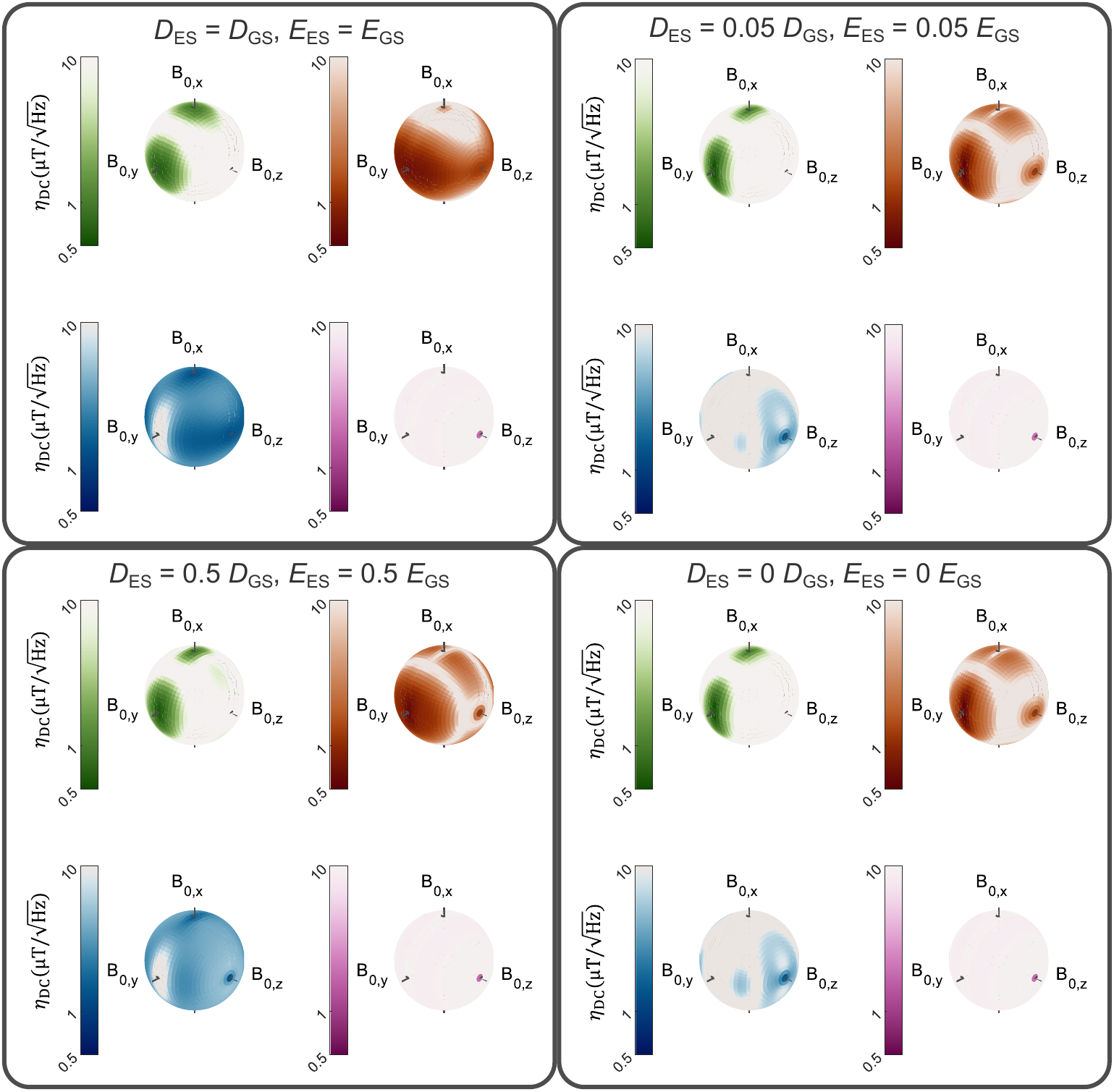}
    \caption{Figure 4 of the main text, calculated for various excited-state zero-field splitting parameters. \label{SIFig:ExcitedState}}
\end{figure}

In our experiments, we do not see spectroscopic signatures of spin transitions in the excited state that would allow us to extract the excited-state zero-field splitting parameters, $D_\text{ES}$, $E_\text{ES}$. In the absence of experimental values, we perform the calculations presented in the main text with the assumption $D_\text{ES} = D_\text{GS}$, $E_\text{ES} = E_\text{GS}$. Figure~\ref{SIFig:ExcitedState} shows that this assumption has little impact for the qualitative findings we present. In this figure, we present the results of Fig.~4a-d of the main text, calculated for different values of $D_\text{ES}$, $E_\text{ES}$. For $D_\text{ES} \neq D_\text{GS}$, $E_\text{ES} \neq E_\text{GS}$, we observe some changes with the occurrence of \textit{blind arcs} where no resonance presents significant contrast when $D_\text{ES} \ll D_\text{GS}$, $E_\text{ES} \ll E_\text{GS}$. Nonetheless, for most bias-field configurations, there is at least one cwODMR resonance that provides significant sensitivity. 

\newpage

\bibliography{myLibrary}